 \documentclass[final,5p,times,twocolumn,authoryear]{elsarticle}

\usepackage{amssymb}
\usepackage{lipsum}
\usepackage{url}

\journal{Results in Physics}

\begin{document}

\begin{frontmatter}



\title{Status Report on Global Pulsar-Timing-Array Efforts to Detect Gravitational Waves}


\author[FSI]{Verbiest, Joris P.~W.} 
\author[UWM]{Vigeland, Sarah J.} 
\author[Milan,Bonn]{Porayko, Nataliya K.} 
\author[SHAO,Kavli]{Chen, Siyuan} 
\author[Swin,Ozgrav]{Reardon, Daniel J.} 
\affiliation[FSI]{organization={Florida Space Institute, University of Central Florida},
            addressline={12354 Research Parkway}, 
            city={Orlando},
            postcode={32826}, 
            state={Florida},
            country={USA}}
\affiliation[UWM]{organization={Center for Gravitation, Cosmology and Astrophysics, Department of
    Physics, University of Wisconsin-Milwaukee},
  addressline={P.O.\ Box 413},
  city={Milwaukee},
  postcode={53201},
  state={Wisconsin},
  country={USA}}
\affiliation[Milan]{organization={Dipartimento di Fisica ``G.\ Occhialini'', Universita degli
    Studi di Milano-Bicocca},
  addressline={Piazza della Scienza 3},
  city={Milano},
  postcode={20126},
  country={Italy}}
\affiliation[Bonn]{organization={Max-Planck-Institut fuer Radioastronomie},
  addressline={Auf dem Huegel 69},
  city={Bonn},
  postcode={53121},
  country={Germany}}
\affiliation[SHAO]{organization={Shanghai Astronomical Observatory, Chinese Academy of Sciences},
  addressline={80 Nandan Road},
  city={Shanghai},
  postcode={200030},
  country={P.~R.~China}}
\affiliation[Kavli]{organization={Kavli Institute for Astronomy and Astrophysics, Peking
    University},
  addressline={5 Yiheyuan Road},
  city={Beijing},
  postcode={100871},
  country={P.~R.~China}}
\affiliation[Swin]{organization={Centre for Astrophysics and Supercomputing, Swinburne University of
    Technology},
  addressline={P.O.\ Box 218},
  city={Hawthorn},
  postcode={VIC 3122},
  country={Australia}}
\affiliation[Ozgrav]{organization={OzGrav: The Australian Research Council Centre of Excellence for
    Gravitational Wave Discovery}, 
city={Hawthorn},
postcode={VIC 3122},
country={Australia}}
  
\begin{abstract}
The stability of the spin of pulsars and the precision with which these spins can be determined,
allows many unique tests of interest to physics and astrophysics. Perhaps the most challenging and
revolutionary of these, is the detection of nanohertz gravitational waves. An increasing number of
efforts to detect and study long-period gravitational waves by timing an array of pulsars have been
ongoing for several decades and the field is moving ever closer to actual gravitational-wave
science. In this review article, we summarise the state of this field by presenting the current
sensitivity to gravitational waves and by reviewing recent progress along the multiple lines of
research that are part of the continuous push towards greater sensitivity. We also briefly review
some of the most recent efforts at astrophysical interpretation of the most recent GW estimates
derived from pulsar timing.
\end{abstract}



\begin{keyword}
Pulsar Timing Array \sep Gravitational Waves \sep IISM \sep Gravitational wave detection

\PACS 04.80.Nn \sep 04.30.-w \sep 97.60.Gb

\end{keyword}

\end{frontmatter}

\tableofcontents



\section{Introduction}\label{sec:intro}

When the cores of moderately massive stars collapse after nuclear fusion ceases, neutron stars can
be formed. These objects conserve a lot of the original angular moment of the progenitor star, but
are far more compact, leading to rapid spins on the order of fractions of a second. A significant
number of these neutron stars furthermore preserve their binary companion stars and can gain further
angular momentum when this companion star evolves into a red giant -- this is the generally agreed
way in which so-called millisecond pulsars (MSPs) are created \citep{lor08}.

Many neutron stars can be detected on Earth due to radiation that is emitted along their magnetic
field lines. While the exact emission mechanism is not fully understood to date \citep{mrm21}, the
``lighthouse model'' of pulsar emission is undisputed. It posits that the magnetic axis of the
neutron star is misaligned with the rotation axis and that consequently the radiation that emanates
from the magnetic pole is swept around in space like the beam from a lighthouse. If the Earth is
fortunate to be positioned within that sweeping cone of emission, the neutron star can be observed
as a series of radiation pulses, in which case the object is typically referred to as a pulsar
\citep{hbp+68}.

Due to the extremely stable spin, the pulsar's sequence of pulses that can be observed on Earth is
typically extremely predictable. While there are a variety of perturbing effects observable in
``slow'' pulsars \citep{vs18}, the MSPs tend to be highly predictable. Consequently, these pulses
can be used to determine a so-called pulsar-timing model which mathematically describes all effects
that impact the arrival time of a pulsar's pulses on Earth. Such effects include obviously the spin
period and the period derivative (due to energy loss) of the pulsar, but also a description of the
binary orbit the pulsar might inhabit (often including relativistic orbital effects), delays
incurred during propagation through the ionized interstellar medium and astrometric properties like
the position, proper motion and distance of the pulsar to Earth. Through fitting of the timing
model to the arrival times of the pulses at the observatory, any and all of the aspects affecting
the timing model can be studied in great detail, see \citet{lk05} for a review.

Probably the most prized phenomenon that could affect the times-of-arrival (ToAs) of pulsar pulses,
is the stretching and squeezing of space-time by gravitational waves (GWs). The fact that these exotic
predictions of general relativity would impact pulsar ToAs has been appreciated since the late 1970s
\citep{saz78,det79}, although in principle it would be hard to differentiate those signals from many
other noise sources in the measurement data (see Section~\ref{ssec:Noise}). \citet{hd83} provided
the solution to this problem, by deriving the mathematical prediction of the \emph{correlated}
signal incurred by ToAs due to GWs. Specifically, \citet{hd83} demonstrated that the GW impact would
display a nearly quadrupolar correlation signature as a function of the angular separation between
the pulsars whose data are being correlated. This idea of hunting for a signal that is not uniquely
identifiable within the data set of a single pulsar, but that can only be detected through timing of
multiple pulsars, gave rise to the idea of the \emph{pulsar timing array (PTA)}, which was first
proposed by \citet{fb90} and \citet{rom89}. Specifically, they proposed that three unique signals
could be pulled from timing an array of pulsars: monopolar correlations would relate to
imperfections in the clock standards used \citep{hgc+20}, dipolar correlations would relate to
errors in the Solar-System ephemerides used \citep{cgl+18,gllc19,vts+20} and finally, quadrupolar
correlations would be caused by gravitational waves.

A number of reviews have been published related to the concept and workings of PTAs. Since the
present article is mostly interested in providing an overview of recent results, we will refer the
interested reader now to review articles that cover other aspects of PTA experiments. The most
comprehensive and recent review is the book by \citet{tay21a}, which covers particularly the theory
and software used for GW detection with PTAs in great detail. A more introductory overview can be found in \citet{vob21} and a somewhat shorter treatment that
includes a review of the state of the art in PTA research nearly a decade ago, is given by
\citet{lom15}. A basic review article that includes an overview of the various software packages
that were used in PTA research half a decade ago, can be found in \citet{hd17}. \citet{tib18}
presents an overview of the PTA concept, with particular focus on various aspects directly
related to pulsar timing. A thorough review of the astrophysical sources of GWs that could be
detectable by PTAs was recently published by \citet{btc+19}, while the theoretical treatment of how the different types of sources impact the timing measurements can be found in \citet{ps18}. Finally, a comprehensive description of the
many factors that drive measurement uncertainty in PTA work was provided by \citet{vs18}. 

In this paper, we will review the recent progress in PTA research. Specifically, in
Section~\ref{sec:PTAs}, we will briefly summarize the various efforts that are being undertaken
around the world to detect GWs in PTA data. Section~\ref{sec:Aspects} will provide an overview of
the aspects that compose the entirety of PTA science; and the state-of-the-art of each of these
aspects will be further described in subsequent sections: existent data sets
(Section~\ref{ssec:Data}), timing software (Section~\ref{ssec:Software}), noise modeling
(Section~\ref{ssec:Noise}) and current challenges including various propagation delays
(Section~\ref{ssec:IISM}).
The current state of published GW results from PTAs is given in
Section~\ref{sec:signals} and in Section~\ref{sec:Astro}, we will briefly summarise the
astrophysical implications of those results. 
In Section~\ref{sec:Future} we will
present a few of the most eagerly anticipated developments in PTA research and finally, in
Section~\ref{sec:Conc} we sum up the most pertinent points of this review. 

%
%
%
%
%

\section{The PTA Landscape}\label{sec:PTAs}
Even though the concept of PTAs has been around since the late 1980s, at that time the practical
implementation was effectively impossible. By 1990, only eight MSPs were known. Four of those MSPs
(PSRs~B1821$-$24A, B1620$-$26, B1516+02A and B1516+02B) inhabited globular clusters -- such pulsars
are randomly accelerated in the highly dense gravitational environment and can therefore not be
timed with high precision. Of the remaining four MSPs, one (PSR~B1957+20) inhabited a so-called
``black-widow'' system in which the ultra-light companion star was believed to be stripped apart,
causing the orbit (and hence the timing) to be highly unstable\footnote{Many decades later it would
be discovered that such black-widow systems can also be stable and some are presently being used in
PTA experiments, see, e.g., \citet{bjs+20,dcl+16}.}. The original MSP, PSR~B1937+21, also did not possess the high timing stability
required for PTA experiments, even though on short time scales (weeks to months) its precision was
phenomenal, due to its high brightness. This left only two MSPs that could realistically be used for
PTA experiments, which was insufficient for allowing the detection of any correlated signal.

A number of highly successful pulsar surveys in the Southern hemisphere changed this situation from
the mid-1990s onwards. The Parkes-70cm survey \citep{jlh+93} discovered 18 new MSPs in the Galactic disk,
the Parkes Multibeam Pulsar Survey \citep{mlc+01} discovered 30 new MSPs and simultaneously, the
Parkes-Swinburne Multibeam Survey \citep{eb01} found another 20 MSPs. In total, by the end of 2005,
no less than 67 MSPs were known in the Milky Way and very soon the first monitoring results
indicated that the timing precision and stability of many of these objects was sufficient to make
PTA research a realistic endeavor \citep{vbc+09}.

The success of the Parkes telescope in these pulsar surveys was primarily due to its opportune
position in the Southern hemisphere, from where the Galactic centre can be seen. Since most pulsars
inhabit the Galactic disk \citep[even if MSPs have a somewhat higher scale height than most heavy
  stars, see][]{lfl+06}, this placed Parkes in a unique position and consequently they were the
first to formally set up a pulsar timing array: the Parkes Pulsar Timing Array or PPTA
\citep{mhb+13}, soon followed by the European Pulsar Timing Array or EPTA \citep{dcl+16} and the
North-American Nanohertz Observatory for Gravitational Waves or NANOGrav \citep{dfg+13}. Even though
all these PTAs were formally started in the first decade of the new millennium, many use data from
pulsar-monitoring campaigns that predate the formal founding of the projects, which means that many
of the PTA data sets date back to the mid-90s or even earlier. Soon after these three original PTAs
commenced operations, the scientific benefit of international collaboration in this high-demands and
high-stakes project was recognised, which led to the founding of the International Pulsar Timing
Array or IPTA. A description of the organizational set-up of the IPTA, especially in those earlier
days, can be found in \citet{man13}, a more comprehensive description of their observing programs
and sensitivity was given in \citet{vlh+16} and updated in \citet{pdd+19}.

In more recent years, new and upgraded telescopes have joined the PTA endeavor. Specifically the
upgraded Giant Metre-Wave Radio Telescope (uGMRT) in India 
forms the Indian Pulsar Timing Array or InPTA \citep{jab+18,jgp+22}, which formally
joined the IPTA in 2022. The Five-hundred metre Aperture Spherical Telescope (FAST) forms the
backbone for the recently established Chinese Pulsar Timing Array \citep[CPTA][]{lee16}, which is
planning to also incorporate the 110-m Qitai Radio Telescope (QTT). The relatively new MeerKAT
telescope array in South Africa also recently commenced its own PTA program, the MeerKAT Pulsar
Timing Array or MPTA \citep{msb+23}. The MPTA and CPTA have not formally joined the IPTA, but they
have been observer members with the intent to formally join at a later date and the MPTA has signed
a data-sharing agreement with the IPTA already. In Russia, data from
the Puschino Radio Astronomy Observatory were used for PTA-like experiments over ten years ago
\citep{rod11}, but no results have been published recently. Finally, upgraded radio telescopes of
the Argentine Institute of Radio Astronomy are also being used for MSP monitoring, with the ultimate
aim to aid PTA studies \citep{glc+20,sdc+21}. A summary of the pulsars being monitored by the
various PTAs can be found in \ref{app:PSRs} and a list of the various observatories
involved in PTA experiments, along with some key properties of their observational data, is included
in \ref{app:obs}.

%

%


\section{Steps in the PTA Process}\label{sec:Aspects}
In essence, there are five steps in the scientific process of PTA research. First, pulsars are
\emph{observed}, typically at radio observatories \citep[but see also][]{faa+22}. The resulting data
are written to disk for off-line processing. In the next step, which could be called ``\emph{basic
timing}'', these observational data get cleaned and reduced in order to ultimately convert the
observations into a set of pulse ToAs and pulsar timing models. These timing models and ToAs are fed
into the third stage (\emph{noise modeling}), where the non-deterministic processes that affect the
timing are determined. In an iterative way, this noise modeling further improves the timing
model. In case of major changes to some of the parameters, some observations may need to be
scrutinized in more detail or some part of the analysis may need to be re-done. The fourth step adds
a further level of complication: in the ``\emph{GW Analysis}'' step, the timing model is expanded to
include correlated signals. At this step the data from all pulsars have to be simultaneously analyzed
in a computationally-intensive Bayesian analysis that attempts to disentangle effects of the
deterministic timing model, pulsar-specific noise parameters and correlated effects from clock
errors, Solar-System ephemeris errors and GWs. Again, some iteration with the previous steps may be
required. Finally, once the correlated signals are determined and the rest of the parameters have
converged onto stable values, the final step of the \emph{astrophysical interpretation} can be
commenced.

In the following few sections, we will describe the various timing data sets that have been
released, list the software packages that are typically used in these analyses, compare the noise
modeling efforts of the major PTAs and discuss some of the more interesting sources of noise. After
that, the
recent GW analysis results and their astrophysical interpretations will be discussed.


\section{PTA Data Sets}\label{ssec:Data}
As discussed above, four PTAs (EPTA, InPTA, NANOGrav and PPTA) have joined forces within the
IPTA. Beyond those the MPTA have released a data set and the CPTA have not yet released their data,
but have published a description of their data set. The six most recently
published data sets from these PTAs will be briefly described below and their main characteristics
will be listed in \ref{app:obs} for the observational characteristics of the different PTAs
and \ref{app:PSRs} for the list of pulsars monitored by the PTAs.


\subsection{EPTA}
The EPTA combines data from the five major radio astronomical telescopes in Europe: the Effelsberg
100-m radio telescope in Germany, the Nan{\c c}ay radio telescope in France, the Lovell telescope at
Jodrell Bank Observatory in the UK, the Westerbork Synthesis Radio Telescope (WSRT) in the
Netherlands and the Sardinia Radio Telescope (SRT) in Italy \citep{dcl+16}. In addition, they
coherently combine signals from these five radio telescopes to synthesize a continent-sized radio
telescope with much higher sensitivity: the Large European Array for Pulsars \citep[LEAP,][]{bjk+16},
which was first included in the most recent analysis \citep{eab+23}. (Note that so far, SRT data
have only been included in EPTA data sets as part of the LEAP data, but not as independent
ToAs. Similarly, WSRT stopped providing independent ToAs in mid 2015 due to significant changes to
the telescope receiver suite, after which it has only provided data that were merged into LEAP
ToAs.) The low-frequency SKA pathfinder LOFAR \citep{vwg+13} commenced a pulsar-monitoring campaign
about a decade ago \citep{kvh+16} and is expected to contribute to EPTA data sets in the near future
\citep{dvt+20}. Similarly, the French LOFAR extension NenuFAR \citep{bgt+21} may contribute at even
lower frequencies. 

The EPTA has previously published data on 42 pulsars \citep{dcl+16}, but their most recent release
only contained the 25 MSPs \citep{eab+23} that were expected to contribute the most GW sensitivity
\citep{spf+23}. The timing programs at some of the EPTA observatories go back into the early 1990s,
providing a full baseline of up to $\sim$25 years, but their most constraining results were derived
from the $\sim$10-year long subset of data obtained with modern recording systems, although
investigations into possible systematic effects or higher-level noise modeling for the earlier data
may yet enable usage of the earlier data to further increase sensitivity. The variety of
observatories contributing to this data set gives rise to a wide diversity of observing
frequencies. The latest EPTA data release contained observations from 323\,MHz at the low end up to
4.8\,GHz at the upper end. Future inclusion of LOFAR data would push the lowest frequency down to
$\sim$100\,MHz (inclusion of NenuFAR would bring the lowest frequency down below 100\,MHz), while
observations at the Effelsberg observatory may push the highest frequency up to 5 or even 9\,GHz in
future releases \citep{lkg+16}.

\subsection{InPTA}
The youngest of the IPTA's consortia published its first data release only recently
\citep{tnr+22}. It contains 14 MSPs and concentrates on low-frequency observations, varying from
300\,MHz up to 1.46\,GHz. The data set contains data from the recently upgraded GMRT and
consequently is only 3.5 years long. Future additions of data from the Ooty Radio Telescope in
Southern India may, however, be possible as well. Due to the low observing frequencies, the GMRT has
concentrated primarily on mitigating interstellar propagation effects \citep[see also
Section~\ref{ssec:IISM}]{jgp+22} and teamed up with the EPTA's higher-frequency data set
in order to derive GW results \citep{eab+23}.

\subsection{NANOGrav}
The NANOGrav PTA has so far been dominated by the Arecibo and Green Bank radio telescopes
\citep{dfg+13}, but the demise of the Arecibo 1000-foot telescope forced this PTA to look for other
powerful telescopes to complement the GBT. In their most recent data release \citep{aaa+23a} they
included pulsar-timing data from the VLA for the first time and a combination with data from the
CHIME telescope is being prepared \citep{cab+21,goo21}.

The most recent data release from NANOGrav \citep{aaa+23a} contains nearly 16 years of data on 68
MSPs and has the widest persistent frequency coverage of any PTA, from 237\,MHz up to 3\,GHz.

\subsection{PPTA}
Formally the oldest PTA, the PPTA is centred around the 64-m Parkes radio telescope, Murriyang. Out of the four IPTA consortia it is the PTA with the smallest collecting area, but also the only one
situated in the Southern hemisphere, providing it with the great advantage of being the only PTA
with access to some excellent MSPs in the Southern sky. The third and latest PPTA data release
\citep{zrk+23} counted 32 MSPs with timing baselines up to 18 years. The PPTA originally used
observations at three observing bands \citep{mhb+13}, with central frequencies around 700\,MHz, 1.4\,GHz, and 3.1\,GHz, but in the last $\sim$3 years started using an ultra-wide-bandwidth observing system that provides an instantaneous bandwidth from 704\,MHz up to 4.032\,GHz \citep{hmd+20}. (Similar ultra-wide-bandwidth systems have been developed for the Effelsberg and Green Bank telescopes but those have not yet been included in
PTA data releases so far.)

\subsection{CPTA}
The CPTA is not presently a member of the IPTA, but they coordinated their latest GW results to be
released at the same time as those of the other PTAs \citep{xcg+23}. While they have not actually
released their data, the data that were used to obtain their GW result were obtained with the FAST
radio telescope over the last 3.5 years over an observing bandwidth from 1\,GHz up to
1.5\,GHz. Their data set contains 57 MSPs. 

\subsection{MPTA}
Finally, the MPTA carries out an MSP monitoring campaign with the SKA prototype telescope
MeerKAT. Their first and latest data release contained 2.5 years of data on 78 MSPs across an
observing bandwidth from 856\,MHz to 1.712\,GHz. The GW results from the MPTA's data have so far not
been released. 


\section{Pulsar Timing Software}\label{ssec:Software}
Observing software is typically specific to the data-taking system, although many recent systems
employ either the \textsc{dspsr} software package \citep{vb11} or a system based on the
\textsc{casper} platform \citep[see, e.g.,][]{drd+08}. The observational data are most commonly
stored in the \textsc{psrfits} format \citep{hvm04} and initial data reduction is typically carried
out with the \textsc{psrchive} package \citep{vdo12}.

A set of recommendations for pulsar timing practice and for meta-data required in ToA files was
presented in the appendix of \citet{vlh+16} and a closer look at the algorithms used for ToA
determination and template creation was recently published by \citet{wsv+22} who experimentally
confirmed some of the earlier recommendations.

For the analysis of the ToAs and the optimization of the timing models, the \textsc{tempo2} package
\citep{hem06} is most commonly used outside North America, while the NANOGrav collaboration uses
both the \textsc{tempo2} and the independent, \textsc{python}-based \textsc{pint} package
\citep{lrd+21}, in order to allow independent verification of the timing results. The mathematical
basis of the timing-model fit is fundamentally the same for the two packages and is detailed
extensively by \citet{ehm06}. 

For the noise modeling the NANOGrav and PPTA collaborations use the
\textsc{enterprise} package \citep{evtb19}, while the EPTA uses both \textsc{enterprise} and the
slightly older \textsc{temponest} package \citep{lah+14}. The GW analysis is carried out with the
\textsc{enterprise} package by all PTAs, although EPTA also uses the yet-unpublished "42"
package\footnote{Available from \url{https://github.com/caballero-astro/fortytwo}.}. The CPTA uses
42 for both noise modeling and GW analysis. Both \textsc{enterprise} and \textsc{temponest} are implementations of a Bayesian analysis code
as described by \citet{lah+14}, but 
also contain modules with implementations of frequentist statistics for the GWB \citep{abc+09,dfg+13,ccs+15,vite18} and individual sources \citep{esc12a}.

Most recently, the \textsc{ceffyl} package \citep{ltv23} has been published, which aims to directly
interpret PTA spectral data in terms of astrophysical models of the GWB in a computationally highly
efficient way. 

%
%

\section{Noise Analysis}\label{ssec:Noise}
As implied in our earlier discussion, there are four different types of signals that affect
pulsar-timing data but are not included in the deterministic pulsar-timing model. Firstly, there is
\emph{white noise} which is expected to be quantified by the ToA uncertainties. However, these
uncertainties are typically underestimated so that correction factors need to be defined
\citep{vlh+16}. Secondly, there is \emph{achromatic red noise}\footnote{The combination ``achromatic
red'' may appear confusing. Note that ``achromatic'' implies that this type of noise has no
dependence on the \emph{observing} frequency, whereas ``red'' refers to the fact that the power spectrum of
this noise has excess power at low \emph{Fourier} frequencies. In other words ``achromatic red noise'' is a
type of noise that varies slowly in time but affects all photons in an identical way.}, which is
commonly also referred to as \emph{``timing noise''}, although the latter is also often used in a
more generic context. Achromatic red noise is usually assumed to be related to intrinsic
instabilities in the evolution of the pulsar's spin period, although no complete model has been
proposed to date. It is known that this red noise in MSPs is typically at a lower amplitude than it
is for slow pulsars \citep{vbc+09,sc10} but the physical, underlying mechanism for this noise is not
clear.

Achromatic red noise was first modeled in pulsar-timing efforts by \citet{vbv+08} and their
frequentist method was further expanded by \citet{chc+11}. In those early experiments the achromatic
red noise was modeled with a broken power law since the assumption was made that the noise itself
had a power-law shape \citep[consistent with the findings in slow pulsars,][]{hlk10} but that at the
lowest frequencies this power would be absorbed by the fit for spin period and spindown. Presently,
NANOGrav \citep{aaa+23b} as well as the EPTA \citep{eia+23} and InPTA \citep{sdk+23} use power-law
models for the achromatic red noise, while the PPTA uses a variety of noise models depending on the
pulsar being studied \citep[see below]{grs+21, rzs+23a}.

The third type of noise affecting pulsar-timing data, is \emph{chromatic red noise}, which refers to
non-deterministic variations in pulse arrival times which have some dependence on the observing
frequency. The key difference with the achromatic red noise is that chromatic red noise could in
principle be uniquely measured (and corrected for), provided observations at different observing
frequencies are available. As discussed in \citet{vs18}, the fractional bandwidth of these
multi-frequency data, as well as the sensitivity in the different bands, is key to such efforts,
implying that ultra-wide observing systems (like the one used by the PPTA) or observations at very
low frequencies (like those provided by the InPTA, LOFAR or CHIME) would provide the best chance at
mitigation of this noise \citep[see also][]{jhm+15}.
Most PTA collaborations use a power-law model for each possible chromatic noise, while NANOGrav
primarily uses the DMX model \citep{leg+18}. Efforts are ongoing to create more custom models for
each pulsar based on the power-law model. Alternatively, as discussed by \citet{jhm+15}, direct DM
measurements at high cadence could also be used to correct higher-frequency data, as for example
suggested by \citet{dvt+20}. Yet another potential approach was put forward by \citet{pdr14}, who
developed software to simultaneously determine a DM value and a ToA from a given wide-band
observation. This method is undergoing testing by NANOGrav, the InPTA and the PPTA. Specifically,
NANOGrav has regularly published both
traditional ``narrow-band'' ToA data sets and such experimental ``wide-band'' timing data
\citep[see, e.g.,][]{aaa+23a}. The InPTA's efforts are described by \citet{pdr+24} and the PPTA's
analysis was recently published by \citet{cpb+23}.

Chromatic red noise is generally caused by
interactions between the radio waves and free electrons in the interstellar medium. Consequently the
main component of such noise is temporal variations in interstellar dispersion, although
higher-order terms caused by interstellar scattering that smears out the pulse profile shape
\citep{lkd+17} or frequency-dependent path-length differences \citep{css16,dvt+19} can also affect
high-precision MSP timing data. A more extensive discussion of the various chromatic noise sources
is presented in Section~\ref{ssec:IISM}. 

The fourth and final type of noise affecting pulsar-timing data are sources of \emph{correlated
noise}, i.e.\ GWs and errors in clock standards and Solar-System ephemerides. These are also
achromatic (i.e.\ independent of observing frequencies) but they have a well-defined correlation
between pulsars, dependent on the angular separation on the sky between the pulsars in question.

\subsection{PPTA}
The most extensive noise models of any PTA were derived by the PPTA \citep{grs+21, rzs+23a}. They
analysed all the types of noise described above, but more importantly, for the achromatic red noise,
they developed different noise models for different observing bands or different observing
systems\footnote{Note this was also done for the first IPTA data release, \citep{vlh+16, lsc+16}.},
as needed and included chromatic red noise with a variable spectral dependence. For ten (out of 26)
pulsars they find that some band or system noise model is favored \citep{grs+21}. Nearly all their
pulsars show evidence for dispersion measure variations (i.e.\ chromatic noise with a square-law
dependence on the observing frequency) and for five pulsars they find chromatic noise with a
different frequency dependence. Time-constrained, correlated and achromatic noise which they ascribe
to temporary pulse-profile shape changes is detected in four pulsars. Such changes (see also
Section~\ref{sssec:events}) have previously been detected in PSR~J1713+0747 \citep{dcl+16,leg+18}
and in PSR~J1643$-$1224 \citep{slk+16}.

%
The more recent PPTA work of \citet{rzs+23a} further investigates the interplay between the noise
models and sensitivity to potential GW signals. Specifically, they demonstrated that the choice of
noise model significantly affects the evidence and recovered spectral properties of a common signal between the pulsars, underscoring the importance of detailed and well-informed noise-modeling in GW detection
efforts.  

\subsection{NANOGrav}
The NANOGrav collaboration's latest noise analysis \citep{aaa+23b} poses a stark contrast to the
work of the PPTA in that they model the red noise only with a simple power-law model and demonstrate
that this is sufficient to model their data in a statistically robust way.
This was enabled using the DMX model \citep{leg+18} to describe the time variation of the dispersion measure as a series of fitted values between subsequent pairs of observations, where each pair consists of two observations within a few days at different radio frequencies. Assuming that the change of the interstellar medium within a few days can be neglected, one can fit for the offset caused by chromatic noise between the two observations in the pair.

They furthermore
demonstrate that the MSPs in their sample can be divided into two groups: one group with strong red
noise that appears to be pulsar-specific and a second group with lower-level red noise that appears
to have identical spectral characteristics between the various pulsars observed \citep[their
Figure 2]{aaa+23b}. PSR~J1939+2134 is the only source that does not match either group, as the
spectral properties of its red noise falls in between the two groups. 

\subsection{InPTA}
With only 3.5 years of data, the InPTA is not expected to be very sensitive to achromatic red noise,
but given their low observing frequencies, variations in interstellar propagation delays
(i.e.\ chromatic red noise) are a prime target of their observing strategy, as already pointed out
in their earlier work on that topic \citep{kmj+21}. The results from their full noise analysis were
presented by \citet{sdk+23} and confirmed that a majority of pulsars (8 out of 14) already showed
significant chromatic noise, while four out of 14 even showed red noise that scaled more strongly
with observing frequency than quadratic, which would be indicative of temporal variations in
interstellar scattering -- a clear demonstration of their high sensitivity to frequency-dependent
propagation effects. Half of their pulsars (7/14) already showed signs of achromatic red noise, at
levels that were consistent with results from the other PTAs.



\subsection{EPTA/InPTA}
The EPTA presented its latest noise analysis in a joint work with the InPTA \citep{eia+23}. They
undertook a comprehensive comparison of the noise properties of various subsets of this combined
data set, namely a) the old EPTA timing data \citep[which effectively constituted the previously
  published results from][]{cll+16}, b) the newer EPTA data and c) the combined EPTA+InPTA data.

The findings were complex and highlighted the challenges of disentangling chromatic and achromatic
red noise in data sets with lacking multi-frequency coverage. Specifically, the newer data set showed
significant complications when compared to the earlier publication, largely due to two
factors. First, the inclusion of the InPTA data significantly enhanced the sensitivity to chromatic
red noise, but only during the few final years during which InPTA data were available; and second,
the extremely long time baseline of the archival EPTA data provided significant sensitivity to any
red-noise process, even if it did not allow discrimination between chromatic or achromatic
noise. The inhomogeneous and time-variable nature of this combined data set provided complex results
regarding the red noise, concluding that nearly a third of the pulsars studied (7 out of 25)
required noise model that were more complex than a simple power law. For example, the red-noise
properties of PSR~J1713+0747 were found to not be stationary \citep[which provides a fascinating
  point of comparison with the results presented by][]{rzs+23a}.
Additionally, tests for chromatic noise similar to \cite{grs+21} yield significant evidence for interstellar scattering in PSR J1600$-$3053.
As a logical follow-up to these
findings, efforts are now underway to combine the decade-long timing data from LOFAR \citep{dvt+20}
with the EPTA/InPTA data, in order to provide longer-term discrimination between different noise
sources.

Unrelated to the main data-analysis efforts, \citet{fss23} also investigated the impact of outlier
ToAs on PTA data analysis and PTA GW results, but concluded any impact would likely be negligible.
%

\subsection{IPTA}
A joint analysis of the data from the major PTAs is still under development, but a direct comparison
of the noise models presented by the various PTAs was already published by
\citet{iaa+23}. Notwithstanding the differences in the approaches of the various PTAs with regards
to noise modeling, it was shown that there were no major inconsistencies between the PTAs' results
and that any differences that did exist in the noise modeling were unlikely to affect searches for
correlated signals between the pulsars of the various data sets.


%

\section{Current Challenges}\label{ssec:IISM}

In addition to red, frequency-dependent noise introduced by time-variable dispersion in the
interstellar medium, a number of other effects can impact pulsar timing precision and have recently
been studied in efforts to improve PTA sensitivity. In this section, we will review three such
effects which are arguably the most relevant ones in present PTA analyses: profile-shape variations
that were for several years considered to be ``dispersion events'' but have recently been shown to
have a different origin, time-variable propagation delays introduced by the Solar Wind and possible
consequences of interstellar scintillation. 

\subsection{Profile-Shape Instabilities}\label{sssec:events}
In recent years, several examples have been uncovered of temporary changes in the shape of MSP pulse
profiles. These findings are of critical importance to any high-precision timing effort since the
stability and reproducibility of the pulse profile is a fundamental assumption underpinning the
technique. Originally, such changes were discovered in the timing of PSR~J1713+0747
\citep{dfg+13,dcl+16,leg+18} and were thought to be due to transient events in the interstellar
medium: minor gas clouds or bubbles that passed through the line of sight and temporarily changed
the group velocity of the radio waves traveling to Earth \citep{lll+21}.

A third such ``event'' that was observed in PSR~J1713+0747 in 2021, however, was monitored more
carefully than the previous two events and it was discovered that the cause of this event was by no
means interstellar. Concerningly, the deviation in timing was shown to be caused by a temporary
change in the pulse-profile shape of the pulsar \citep{ssj+21,lam21a,jcc+22}. A similar change was
earlier seen in PSR~J1643$-$1224 \citep{slk+16} and the PPTA claims to have detected similar
behaviour in PSRs~J0437$-$4715 and J2145$-$0750 as well. These discomforting discoveries have raised
awareness of the need to be on the look-out for instabilities in the pulse profile shapes, which
has so far not been done routinely at any observatory.

The fact that these events affect the timing results is clear, but their impact on GW sensitivity is
so far ill-defined. It is of note, however, that in their 12.5-yr analysis, NANOGrav pointed to
these temporary events as one possible reason why PSR~J1713+0747 was shown to contribute less than
any other pulsar in their data set, to a red-noise signal of interest \citep{abb+20a}.

%

\subsection{Solar Wind}\label{sssec:SW}
The Solar wind is a turbulent plasma that has a complex and rapidly varying distribution
\citep{man12}. Furthermore, the lines of sight to pulsars pass through different parts of the
heliosphere at different times of the year, causing the dispersive impacts on the pulse times of
arrival to be highly variable throughout the year. In terms of how such variations impact PTA
science, it is critical to bear in mind the difference between the additional delays experienced by
the entire band (which is the actual impact on the timing and GW analyses) and the \emph{difference}
in dispersive delay measured across the observing band, as discussed in \citet{vs18} (see, e.g.,
their Figure~6).

For relatively high-frequency observations, or for observations with relatively limited fractional
bandwidth, the differential delays \emph{across} the band may be hard to determine, but the time
variations affecting the entire band may still be highly significant -- even if they cannot
undeniably be identified as time-variable dispersive effects. As a consequence, studies of the
Solar-wind impact on pulsar timing were hampered by the limited sensitivity of early data
sets.

\citet{yhc+07b} were the first to attempt to determine the Solar-wind impact on PTA data and
it was based on their partial success that subsequent PTA analyses \citep[most explicitly][]{vlh+16}
decided to remove all observations within five to ten degrees from the Sun, given the high
likelihood of uncorrectable Solar-wind impacts. A dedicated analysis of the Solar wind effects on
NANOGrav data was presented by \citet{mca+19}, who did detect the Solar wind in some of their
pulsars, but did not see any evidence of variations in the Solar-wind density on timescales of
$\sim11$~years, i.e.~with the Solar cycle. These results stand in contrast with those from
\citet{tsb+21}, who used low-frequency LOFAR data to analyze the impact of Solar-wind variations in
pulsar timing efforts and found clear evidence of Solar-cycle variability in the amplitude of the
Solar-wind-induced delays, even if the significance of these variations was strongly dependent on
the Solar latitude.

In a related publication, \citet{tvs+19} investigated how well analytic models were able to model
the impact of the Solar wind on pulsar-timing data and found that none of the models commonly used
by PTAs were sufficient to mitigate Solar-wind effects at levels required for GW science and that
the remaining Solar-wind impact was significant far beyond the 5-10 degrees commonly
excluded. Specifically, the authors argued that, unless significant progress is made in modeling
Solar-wind impacts on pulsar timing data, all data within 20 (and in some cases even 40) degrees
from the Sun may have to be considered corrupted.

Efforts to improve the negative impact of the Solar wind on PTA data are ongoing along three lines
of research. First, phenomenological modeling of the Solar wind as part of the standard noise
modeling has recently been developed and is starting to be used \citep{sct+24}. Second,
combinations of PTA data with low-frequency data are being used to attempt a more precise
modeling of both Solar-wind and interstellar propagation effects. Recent work by \citet{nkt+24}
demonstrates that the traditional, high-frequency, PTA data are significantly affected by Solar-wind
effects without being able to mitigate these effects, but that combination with low-frequency LOFAR
data resolves this problem. Finally, interdisciplinary efforts at modeling the heliospheric electron
density and space weather based on diverse data sets, to serve the dual purpose of solar
astrophysics and PTA corrections for propagation effects, is being undertaken and has shown some
promising preliminary results \citep{bfm+22,tjc+23}.

A less common impact from the Solar wind on pulsar timing would be the occurrence of coronal mass
ejections (CMEs). These rapidly moving and relatively spatially constrained bullets of plasma occur
regularly in the Solar System, particularly near Solar maximum. The chances of any single
observation being affected by such an event are relatively minor, but given the size of PTA data
sets and the fact that it can be extremely challenging to identify a single ToA that is offset, make
it important to verify if no CMEs are affecting the PTA data. So far, only the InPTA has provided
conclusive evidence of being affected by a CME \citep{kmj+21}, although it is likely other data have
merely not been identified as being affected. Automated software to search for coincidences between
PTA observations and CMEs has been made available in recent years by \citet{stz20}. 
%
%

\subsection{Interstellar Scintillation}\label{sssec:ISS}
The ionized plasma in interstellar space does not merely affect the light propagation time, but
causes a host of more subtle effects as well. A group of related effects are often referred to as
``scattering'' or ``scintillation'', which depending on the context can be used interchangeably or
as distinct observable effects. The physical mechanism that lies at the basis of both scattering and
scintillation, is the fact that individual photons will take slightly different paths through
interstellar space. Due to small-scale differences in the spatial distribution of free electrons,
these different photons will arrive at the observer with slightly different phases, which causes
interference and hence variations in observed brightness. Described in this way, the phenomenon is
typically referred to as \emph{scintillation} and it primarily affects the brightness of the pulsar
observation \citep[although higher-order effects are possible, see][]{vs18}.

A different, but related, situation occurs when beams of radio waves are refracted in ways that
cause their path-length differences to be significantly longer compared to waves that come along the
shortest route. This would cause a significant spread in the arrival times of photons, depending on
the path they traveled through interstellar space. The path-length differences are too significant
to result in effective interference, but instead, the shape of the pulse profile will be smeared
out, as a certain fraction of photons arrives with an appreciable delay. This phenomenon is often
referred to as \emph{scattering}. In itself such smearing of the pulse profile would only worsen
timing precision, as it smooths over sharp features that enhance ToA precision, but in case the
scattering is time-variable, this effect can also introduce significant noise in the timing
data. Specifically, since the refraction angles are strongly frequency-dependent, such time-variable
scattering effects would effectively introduce a chromatic type of red noise, which does not scale
with the square of the observing frequency (like the dispersive propagation delays), but rather be
an even stronger function of observing frequency.

For the following text, we will adhere to the definitions given above, namely that scintillation
refers to minor refractive phase shifts that cause intensity interference at the observer's plane
while scattering refers to larger-scale path-length variations that cause distortions in the pulse
profile. However, the reader be warned that since both effects are fundamentally caused by
refraction of the same interstellar, ionized plasma, many authors habitually refer to both effects
as scattering.

Since scintillation does not in itself affect the pulse arrival time at a level that is commonly
detectable at the timing precision that is typical of PTA data currently, it has so far not been
considered a major complication for PTA research\footnote{In combination with profile evolution and
excessively large ToA bandwidths, scintillation \emph{can} cause issues, although in practice these
are readily avoided, as described by \citet{vs18}.}. It has been used, however, as a complementary
observable that can aid in defining accurate pulsar timing models for binary pulsars. This work was
originally pioneered by \citet{obv02} on PSR~J1141$-$6545, but their analysis was corrected by
\citet{rch+19} who later also applied it to the PPTA pulsar J0437$-$4715 \citep{rcb+20}. Similar
work was done with PTA data specifically by \citet{fpe+16}, \citet{pdd+19}, \citet{wrts22},
\citet{ars23}, \citet{lmv+23}, and \citet{mac+23}. Of more direct relevance to PTA work, is the
claim that scintillation information can be used to constrain (and potentially derive) variations in
the density of free electrons, which causes the dispersive delays in the data. Initial results on
this concept were recently published by \citet{rc23} but have not yet been developed to the point
where they can be practically applied to PTA data.

Scattering as defined above is of more direct consequence to PTA efforts since it directly affects
the timing precision and if it is time-variable, it will be an additional source of red, chromatic,
noise as well. To account for this in a phenomenological way, several PTAs have expanded their
Bayesian noise-modeling efforts to include a chromatic red noise with variable power-law dependence
on the observing frequency. Theoretically the strength of scattering is expected to scale with the
observing frequency as $f^{-4.4}$, based on the assumption of thin-screen Kolmogorov turbulence
\citep{cr98}. Observational studies, however, have shown that this frequency dependence is in
practice often a lot shallower \citep{bcc+04} and can even be time-variable
\citep{lvm+22,btsd19}. Furthermore, the frequency dependence of the scattering does not necessarily
equate to the frequency dependence of the impact on the ToAs, since the way the ToAs are affected is
most strongly influenced by the overall shape; and depending on specific features of the pulse
profile, the impact on the ToAs may not scale as expected. Finally, the effective impact of the
scattering on pulse profile shapes is usually assumed to be a convolution of the pulse profile with
an exponential broadening function, based on the assumption of the scattering happening in an
infinite, thin screen. This assumption has been shown to break down in a number of cases, either due
to the limited size of the scattering screen \citep{wcv+23} or due to high anisotropy in the
scattering medium \citep{gkk+17}. Both the PPTA \citep{grs+21} and the InPTA \citep{sdk+23} have
reported significant scattering variations in several pulsars; the EPTA has reported long-term
scattering variations only in PSR~J1600$-$3053 and NANOGrav did not see any evidence for scattering
variations at all to date \citep{aaa+23b}.

In principle the effects of scattering can be corrected through a process called \emph{cyclic
spectroscopy} \citep{dem11}, but this processing-intensive procedure must be ran on the raw data
stream as it comes off the telescope and can hence not be applied retroactively on archival
data. Some effort has gone into quantifying the potential benefit of such analyses, largely based on
simulations. These investigations have demonstrated that for distant and strongly scattered
pulsars, cyclic spectroscopy could significantly improve the timing quality and hence enlarge the
number of MSPs that could meaningfully contribute to PTA efforts \citep{dsj+21}. 

A final effect from the ionized component of the interstellar medium is effectively a combination of
the effects we called scattering and scintillation earlier. As a basic explanation, consider two ray
bundles that travel from the pulsar to the telescope, one along a mostly straight line and the other
along a significantly longer path due to it being strongly refracted at one location along its travel
path. Within each bundle of rays small-scale differences in phases will cause scintillation, but due
to the long path difference between the two bundles, this scintillation pattern is modulated with a
larger-scale interference effect, namely the interference between the two bundles, which works on
larger scales.

In the frequency-time plane where scintillation is observable
as random patches of intensity variations, this higher-order effect is seen as a corrugated
modulation on top of those random patches. Taking the two-dimensional Fourier Transform of the
intensities in the frequency-time plane returns the so-called \emph{secondary spectrum} in which
this modulated interference can readily be identified as parabolic distributions of power. This
effect was first identified by \citet{smc+01} and it highlights the highly anisotropic nature of the
ionized medium discussed above. In principle these \emph{scintillation arcs} can be related to
well-defined ionized structures in the ionized medium \citep[see, e.g.][]{lmv+23}, but in practice
this may be sufficiently challenging \citep{occ+23} that using these structures for PTA corrections
of interstellar propagation delays is still well beyond current capabilities. This phenomenon
primarily gained interest after \citet{hs08} demonstrated that in some cases the refracted power from
the secondary ray bundle could be sufficient to significantly affect the arrival times measured from
the observed pulses. Consequently, a number of recent campaigns have tried to take stock of the
prevalence of these arcs \citep{wvm+22,mpj+23,occ+23}, but so far no significant effect on timing
has been demonstrated beyond the work of \citet{hs08}.

%



\section{Signals and Detections}\label{sec:signals}
PTAs are expected to detect mostly persistent signals 
in the form of a GW background and 
a number of individual binaries that are  slowly evolving \citep{rsg15,kbh+18}. 
As a consequence, the GW
signal builds up over time and with ever increasing instrumental sensitivity and longer observing
timelines, a detection will slowly arise. This stands in sharp contrast to the GW detections made to
date by the LIGO-Virgo interferometers, which have so far all been 
binary mergers that last much
shorter than the observational timespan \citep{klv18}. This causes a GW detection with PTAs to
consist of three different stages: initially no signal will be detectable, so only limits can be
placed on the amplitude of any GWs present in the data. At some point, these limits will saturate
as the GWs start to significantly affect the timing of the most precisely timed pulsars. The first
real (but ambiguous) signal that can be expected from GWs, is a ubiquitous red noise that equally
affects all pulsars and that has common spectral characteristics in all pulsars -- this type of
signal is called ``common red noise'' (CRN) and would be the first indication of the presence of GWs \citep{rhsa21}. A detection of CRN in itself is, however, not equal to a detection of GWs since the
red noise could have a variety of origins and the consistency between pulsars may be coincidence or
physical, but need not be due to GWs \citep{thk+16,gts+22}. Therefore, the true detection of GWs can only occur when it is
shown that the CRN is correlated between the pulsars in a quadrupolar way \citep{hd83}. A detailed
list of formal requirements to claim a statistically significant detection of GWs in PTA data was
recently presented by \citet{adg+23}. Most importantly, they propose a 5-$\sigma$ minimum
statistical significance in order to differentiate a reliable detection from a chance alignment of
noise. 

\subsection{Limits}\label{ssec:Limits}
The first limit on the amplitude of the Gravitational-Wave Background (GWB) after the formal
commencement of PTA science was placed by the PPTA \citep{jhv+06} and was about an order of
magnitude higher than the most likely expected amplitude of an astrophysical GWB at the time
\citep{jb03,shmv04}. Over the course of the decade that followed, PTA limits on the GWB amplitude
got increasingly more constraining, down to the point where they were providing meaningful
constraints on galaxy-evolution theory \citep{srl+15}. The most constraining limit to date is the
one by \citet{srl+15}, which places a 95\%-confidence upper limit on the dimensionless amplitude of
the GWB of $A_{\rm 1\,yr} < 1.0\times 10^{-15}$ at a reference frequency of 1\,yr$^{-1}$, but some
technical issues (detailed below) rendered this limit unreliable, so it was replaced with the
slightly higher $A_{\rm 1\,yr} < 1.2\times 10^{-15}$ following a recent reanalysis
\citep{rzs+23b}. This limit is closely followed by the most constraining limits from NANOGrav
\citep[$A_{\rm 1\,yr} < 1.45\times 10^{-15}$,][]{abb+18a}, IPTA \citep[$A_{\rm 1\,yr} < 1.7\times
  10^{-15}$,][]{vlh+16} and EPTA \citep[$A_{\rm 1\,yr} < 3.0\times 10^{-15}$,][]{ltm+15}. EPTA and
PPTA did not publish a new limit after 2015, the IPTA only published a limit in 2016 and NANOGrav's
limit barely changed between 2016 and their last limit in 2018. The lack of newly published limits
was due to challenges the PTAs encountered in the form of red noise which saturated their ability to
constrain the GWB further. This spurred an extensive push towards further improvements in data
analysis and developments in instrumental sensitivity and indicated a potential move towards an
actual GWB detection soon.

During this time, there was significant interest in how errors in the Solar System ephemeris (SSE)
would affect PTAs \citep{thk+16}. PTAs can be used to measure the masses of the planets
\citep{chm+10} and search for unmodeled objects in the Solar System \citep{glc18,cgl+18}. It was
also noticed that PTA upper limits on the GWB amplitude were sensitive to the choice of SSE model
used. Specifically, \citet{abb+18a} demonstrated that the DE421 SSE model allowed much more
stringent limits to be placed on the GWB amplitude than more recent, more accurate SSE
models. Consequently, they proposed that some marginalization over some of the more ill-constrained
parameters of the SSE model should be a key part of any PTA GW analysis. By adding those additional
parameters to the analysis, limits would become more conservative and consequently more robust
\citep[see also the more detailed analyses by][]{vts+20,tin18,roe19}. This joint analysis of GW
signal and SSE model would allow a separation of the two signals and would therefore not only
provide more robust limits, but also prevent false GW detections, as was pointed out earlier by
\citet{thk+16}. In practice, the corruptions of the DE421 ephemerides were largely alleviated by
the inclusion of data from the Juno mission, which provided more accurate measurements of Jupiter's
orbit \citep{pfwb21}.

%
In terms of interpreting upper limits on the GWB amplitude, 
\citet{hssa20} showed that Bayesian upper limits are sensitive to the choice of prior on the pulsar red noise due to the covariance between the two. 
Another potential issue with early upper limits is the number of pulsars used: 
\citet{jvst22} showed through simulations that limits on the GWB may be biased if a small number of
pulsars are used. This is an important factor to consider when interpreting early PTA upper limits
since early PTA data sets included significantly fewer pulsars than are used currently. Due to these
issues, some published limits on the GWB amplitude are likely unreliable and lower than they really
should be. 

%

\begin{figure}
  \centering
  \includegraphics[width=0.5\textwidth,angle=0]{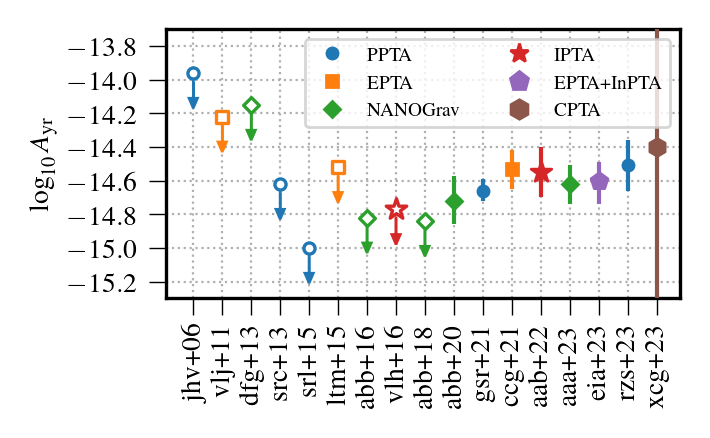}
  \caption{Evolution of PTA sensitivity to gravitational waves. Shown are upper limits on the GW
    amplitude (unfilled markers with arrows) along with more recent measurements of CRN (first four
    filled markers with error bars from the left) and measurements of red noise with quadrupolar
    correlation between the pulsars (final four filled markers with error bars).  Error bars are
    taken from the respective papers: NANOGrav, the EPTA, and the IPTA reported 90\% confidence
    intervals while the PPTA reported 68\% confidence intervals.
    Note that, due to reasons laid out in Section~\ref{ssec:Limits}, some of the limits are
    unreliable and more constrictive than they really should be. The references given on the X-axis
    are:
    \citet{jhv+06,vlj+11,dfg+13,src+13,srl+15,ltm+15,abb+16,vlh+16,abb+18a,abb+20a,gsr+21,ccg+21,aab+22,aaa+23c,eia+23a,rzs+23b,xcg+23}.}
  \label{fig:Limits}
\end{figure}

\subsection{Common Red Noise}\label{ssec:CRN}
Given the saturation of PTA limits, it was already clear that red noise was showing up. Whether this
red noise had identical characteristics between the various pulsars was not clear until the work by
\citet{abb+20a}, who published the first claim of ``Common Red Noise'' in their data, adding that
this might be a suggestive first hint at actual GWs in their data. This work was soon followed up by
confirmatory studies by PPTA \citep{gsr+21}, EPTA \citep{ccg+21} and IPTA \citep{aab+22}, all of
whom confirmed a consistent signal with that presented by NANOGrav, but none of whom found
significant evidence for interpulsar correlations. These results are summarized in
Table~\ref{tab:crn_results}.

Extensive simulations were carried out to verify to what degree this CRN was a cause for
excitement. Specifically, \citet{gsr+21} did report on CRN in the PPTA data set, but also commented
that this noise could be produced by noise inherent to the different pulsars and that there is evidence
that it has to arise from a common process. This work was continued by \citet{zhs+22} who carried
out extensive simulations to show that CRN could easily be the result of a combination of
pulsar-intrinsic timing noise with bad or inaccurate Bayesian priors. Nevertheless, \citet{gts+22}
devised a method to differentiate between truly common red noise and merely 'similar' red noise,
while finally proving that the red noise in the PPTA data was more likely to be truly CRN than
merely a misidentification of spin noise.
Similarly, the EPTA published an extensive analysis of their CRN signal \citep{cbp+22},
demonstrating that the characteristics of the signal were robust to significant changes in the noise
modeling.

\begin{table}
  \begin{tabular}{l | l l}
    \hline
    PTA      & $A_{\rm CRN}/10^{-15}$ & Reference \\
    \hline
    NANOGrav & $1.92^{+0.75}_{-0.55}$ & \citet{abb+20a}\\
    PPTA$^{\dagger}$ & $2.2^{+0.4}_{-0.3}$    & \citet{gsr+21}\\
    EPTA     & $2.95^{+0.9}_{-0.7}$   & \citet{ccg+21}\\
    IPTA     & $2.8^{+1.2}_{-0.8}$    & \citet{aab+22}\\
    \hline
    \end{tabular}
  \caption{Summary of the amplitudes of CRN detections from the various PTAs, under the assumption
    of a spectral index of $\alpha = -2/3$, as expected from a GWB formed from SMBHB
    mergers. $^{\dagger}$ The PPTA reported a 1-$\sigma$ uncertainty, whereas the other PTAs all
    reported 95\% confidence intervals.}
  \label{tab:crn_results}
  \end{table}

%
%
%
%

\subsection{Hellings-Downs Correlation}\label{ssec:HD}

The search for the tell-tale quadrupolar correlation signal in PTA data is more complex than it may
seem on the surface. As early as 2016, \citet{thk+16} warned of the potential corruption of
different types of correlations: monopolar correlations introduced by clock errors and
pseudo-dipolar correlations\footnote{Correlations introduced by errors in the Solar-System
ephemerides are strictly dipolar at any given point in time, but the direction of the dipole slowly
moves on the sky, causing the actual effect in a long-term data set to be more complex than a
straightforward dipolar correlation.} caused by Solar-System ephemeris errors can readily be
conflated in analyses that don't explicitly attempt to differentiate between the three types of
correlations. Furthermore, 
the natural variance of the HD curve, 
and the use of a finite number of pulsars 
to measure the correlations 
complicate searches for the GWB 
\citep{bn22,all23,ar23}. A
useful and accessible discussion of a variety of tricky aspects related to the quadrupolar
correlation sought for in PTA data can be found in \citet{ra23}.
%

\begin{figure}
  \centering
  \includegraphics[width=0.5\textwidth,angle=0]{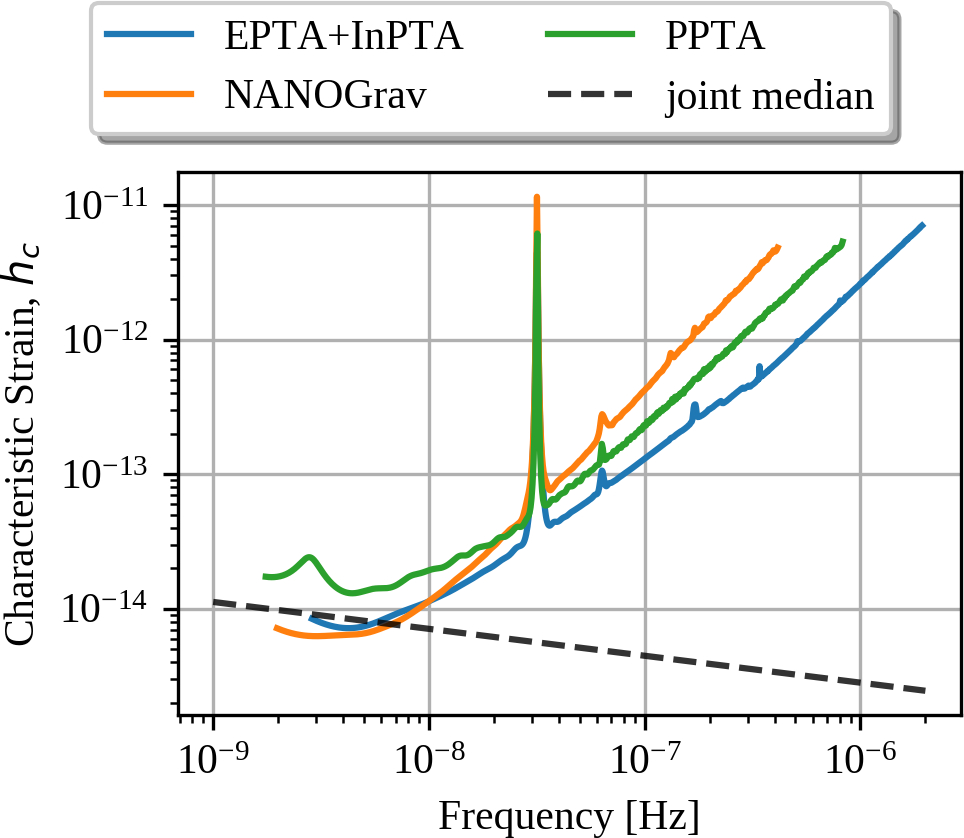}
  \caption{Sensitivity to a GWB for the four main PTA collaborations. Shown are the sensitivity
    curves for the most recently published GWB analyses of the four main PTA collaborations
    \citep{eia+23a,aaa+23c,rzs+23b}, where EPTA and InPTA provided a joint analysis. The dashed line
    shows the best-fit GWB spectrum based on the joint probability distribution of the three
    publications. Reproduced from \citet{iaa+23}.}
  \label{fig:Sensitivity}
\end{figure}

With those caveats, 
PTAs searched their latest data sets and found for the first time evidence for correlations,
building on the detections of CRN found in previous data sets that were discussed in the previous
section.  These results were published in a coordinated fashion on 29 June 2023 by the four PTAs
that constitute the IPTA as well as the CPTA. Results are summarized in Table~\ref{tab:gwb_results}
and the relevant sensitivity curves are reproduced in Figure~\ref{fig:Sensitivity}. These results
show that all the major PTAs have consistent results, albeit with variable sensitivity. NANOGrav
obtained a 3-4-$\sigma$ significance of a quadrupole-correlated signal \citep{aaa+23c}, the EPTA
achieved a $\sim$3-$\sigma$ significance based on their most recent data combined with the InPTA's
first data release \citep{eia+23a}. The PPTA obtained a $\sim$2-$\sigma$ result but note that the
signal they detect is not stationary: in the earlier part of their data the amplitude they detect is
different \citep{rzs+23b}. Finally, the CPTA presented a $\sim$4.6-$\sigma$ result for GWs at the
frequency of 14\,nHz \citep{xcg+23}, but since their analysis is fundamentally different from the analyses of the
other PTAs; and since their data have so far not been publicly released (in contrast to the data
from the other PTAs), it is presently impossible to meaningfully compare this value to the levels of
detection significance from the other PTAs.
\begin{table}
\begin{tabular}{l | c c l}
 \hline
 PTA        & $\log_{10} A_\mathrm{yr}$ & $\gamma$ & Reference\\ 
 \hline
 EPTA+InPTA$^*$ & $-14.10^{+0.25}_{-0.44}$  & $3.03^{+1.02}_{-0.67}$ & eia+23b\\ 
 NANOGrav   & $-14.19^{+0.22}_{-0.24}$  & $3.2\pm0.6$ & aaa+23c\\ 
 CPTA$^\dagger$       & $-14.4^{+1.0}_{-2.8}$     & $[0, 6.6]$ & xcg+23\\ 
 PPTA       & $-14.50^{+0.14}_{-0.16}$  & $3.87 \pm 0.36$ & rzs+23b\\ 
 \hline
\end{tabular}
\caption{Summary of GWB results from the PPTA, EPTA+InPTA, CPTA, and NANOGrav. The GWB can be
  described in terms of the characteristic strain spectrum $h_c(f) = A_\mathrm{yr}
  (f/f_\mathrm{yr})^\alpha$ or the residual power spectral density $P(f) = h_c(f)^2/(12\pi^2f^3) =
  A_\mathrm{yr}^2/(12\pi^2) (f/f_\mathrm{yr})^{-\gamma} \mathrm{yr}^{-3}$, where
  $\gamma=3-2\alpha$. Quoted are 90\% credible intervals, except for the PPTA, where the 68\%
  confidence interval is given. $^*$For the EPTA/InPTA combination the results from their DR2new+
  data set are shown. 
  $^\dagger$Note
  the CPTA reported their results as constraints on the characteristic strain spectrum, with
  $\alpha\in[-1.8, 1.5]$. The references given, are: \citet{eia+23a}, \citet{aaa+23c},
  \citet{xcg+23} and \citet{rzs+23b}.  }
\label{tab:gwb_results}
\end{table}

None of the PTAs reached the 5-$\sigma$ level of significance required
according to \citet{adg+23}, so that formally none of the PTAs have been able to claim a
statistically significant \emph{detection} of GWs in the nHz band, meaning that the measured
correlations can merely be seen as \emph{evidence for} GWs in the PTA data. Hopes are that the
addition of more recent data and the combination of data from all the various PTAs would push the
significance levels to beyond the 5-$\sigma$ level, although these efforts are still ongoing. Also
the high sensitivity of the CPTA data, following a very brief observing campaign, raises hopes for
the MPTA data set which has yet to announce its GW results.

\subsection{Individual Supermassive Binary Black Holes}\label{sec:CW}
While the preceding sections focused on a broadband spectrum of GWs produced by a stochastic GWB, an
alternative source of GWs would be individual supermassive binary black holes which would emit GWs at a single
frequency provided they are in circular orbits that do not evolve over the time span of the
observational campaign. 

\citet{psb+18} used a dedicated high-cadence data set on PSR~J1713+0747 to
constrain single sources of GWs at the highest GW frequencies accessible to PTAs, deriving $A_{\rm
  1\,\mu Hz} < 3.5\times 10^{-13}$ for the sky-averaged limit on the amplitude of individual GW
sources at a GW frequency of 1\,$\mu$Hz, which equates to periods of about two weeks. 
Efforts have also been undertaken to use PTA data sets to detect GWs from a single binary. 
\citet{fbb+23} analysed the second data release of the IPTA \citep{pdd+19} for continuous GWs and
derived a 95\% upper limit on the gravitational strain of these binaries between $10^{-13}$ and
$10^{-14}$, depending on the exact frequency, with optimal frequency around 10\,nHz. No significant
signals were found.

More recently, searches were undertaken by the 
EPTA/InPTA \citep{aaa+23h} and NANOGrav \citep{aaa+23e} 
collaborations. EPTA/InPTA and NANOGrav both found weak evidence for a signal near 4\,nHz, but both
studies concluded this signal was lacking statistical significance and did not provide sufficient
evidence for quadrupolar correlations. \citet{aaa+23h} concluded their 4-5\,nHz ``signal'' could be
a misinterpretation of a GWB signal and pointed out data sets with more sensitivity would be needed
to enable a decisive conclusion. \citet{aaa+23e} furthermore found some weak evidence for a feature
at 170\,nHz but decided this was also unlikely to be a real GW source given the high GW frequency
and the fact that the signal seemed dominated by a single pulsar which has an orbit with a similar
period as that of the candidate GW signal.

While the majority of searches for GWs from individual binaries have assumed circular orbits, some binaries emitting GWs at nanohertz frequencies may have significant eccentricities. These searches are more computationally intensive and require more complex signal models \citep{thgm16,sght20,sus23}. 
\citet{aab+23}
used the most recent NANOGrav data set to perform a search for GWs from an eccentric binary at a particular SMBHB candidate, 3C66B 
\citep{simt03}. They derived an upper limit on the chirp mass of $(1.98\pm0.05)\times 10^9\,{\rm
  M}_{\odot}$ (95\% confidence) for orbital eccentricities lower than 0.5 and a symmetric mass ratio
above 0.1.

To conclude our discussion of individual SMBHB sources, we point out that many of the potential
pitfalls that made early GWB limits unreliable (such as inappropriate priors, inaccurate SSE
models, possible bias in pulsar selections and excessive or lacking modeling of intrinsic pulsar noise)
can also affect limits on individual GW sources. Increasing awareness does imply any such effects
are likely to be less prominent in more recent analyses, however.


%
%


\subsection{Anisotropy in the Background}

In between the extreme scenarios of a GWB and an individual GW source, is an anisotropic background
of GWs. This is essentially a combination of a GWB with several bright GW sources that stand out
beyond the background. In such a scenario the overall anisotropic character of the GWB would be
identifiable before any individual source would be confidently detectable \citep[see,
  e.g.,][]{ses13b,gklm23}. According to the work by \citet{ptr22}, a 20-year data set with
characteristics similar to those of the NANOGrav PTA, would have a realistic chance of detecting
anistropy at the 3-$\sigma$ level. This work is based on the assumption that the GW signal strength
scales with $\sqrt{N_{\rm psr}\left(N_{\rm psr}-1\right)/2}$ (where $N_{\rm psr}$ is the
number of pulsars in the array), which explains why NANOGrav is predicted to have the best chance of
detecting such a GWB, given that it has the largest number of pulsars in its array. However, this
scaling law \citep[taken from][]{sejr13} implicitly assumes all pulsars are in the ``strong-signal''
regime and contribute equally -- which is likely not the case in any realistic PTA. Also, this
predictive work assumed an anisotropy level\footnote{The anisotropy level is quantified as a ratio
of the squared angular power $C$ at angles of 180$^{\circ}$ ($l=1$) to the squared angular power in
the multipole mode ($l=0$). A fully isotropic background would only have power in the multipole
mode, reducing the ratio to 0, whereas an anisotropic background would have power at the higher modes
as well, creating a positive value for the ratio $C_{\rm l=1}/C_{\rm l=0}$.} of $C_{\rm l=1}/C_{\rm
  l=0} = 30\%$, but the latest work by \citet{aaa+23f} already constrained the GWB anisotropy below
this level: $C_{\rm l=1}/C_{\rm l=0} < 27\%$.


%
%
%

\section{Astrophysical Interpretation}\label{sec:Astro}

Generally the most likely source of GWs in the PTA band is considered to be supermassive black-hole
binaries (SMBHBs) that are the consequence of hierarchical galaxy evolution. In this case, two
different types of GWs could be detectable by PTAs: the purely stochastic GWB that would be
characterized by a red-noise spectrum, or individual SMBHB sources that would be monochromatic and
unevolving within our lifetime. It is considered increasingly likely that a combination of these
sources would be detected -- or an early indication of such a combination, i.e.\ a GWB with
anisotropic features. Notwithstanding the focus that has been placed on SMBHBs, a diverse range of
alternative sources of GWs has also been proposed, most commonly related to strings or
inflation. Alternatively, the PTA data can also be used to detect or place constraints on GW
polarization or on ultra-light dark matter (ULDM), which is expected to result in monopolar rather
than quadrupolar correlations.

In this section, we will briefly review some of the main research that has been recently derived on
these sources based on the most recent PTA results. 

\subsection{Supermassive Binary Black Holes}
A good review on the basics of SMBHBs and how they can be detected through GWs, is given by
\citet{dc23}. Fundamentally, as galaxies merge, the supermassive black holes at their centers form
binaries that ultimately lead to mergers. Such a SMBHB emits gravitational waves with a GW frequency
twice the orbital frequency and can hence be detectable in PTA data for orbital periods in the range
of years to decades. Since most of these binaries are expected to be relatively distant and faint,
they may not be easily detected individually, but their combined stochastic sum is expected to reach
a level that should soon be detectable by PTAs.

Based on the signals reported in Section~\ref{ssec:HD}, both NANOGrav \citep{aaa+23d} and EPTA/InPTA
\citep{aaa+23i} have studied the implications of this signal for SMBHB models. Both groups found
that many different models for galaxy evolution and SMBHB formation and evolution are technically
statistically consistent with the data, but that typically either many parameters are on the edge of
their predicted ranges, or that a few parameters are significantly inconsistent with
expectations. Specifically, \citet{aaa+23i} point out -- based on the results from L-galaxies simulations \citep{hwt+15,isbc22} -- that the high amplitude of the signal implies
a shorter SMBHB evolutionary timescale and, more importantly, a higher gas accretion rate than usually expected, but that this may cause tension with
other observables like the AGN luminosity function. Furthermore, the fact that the spectral slope is
shallower than expected may be interpreted as an indication that the GWB is not dominated by
circular SMBHBs, but that eccentric SMBHBs or other environmental effects play a significant role as
well, or that small-number statistics may be at play (in which case a detection of anisotropy in the
background should be expected in the near future).

The methodology behind these analyses was studied by \citet{vsss23} who found that the results were
mostly robust, provided care was taken with the assumed GWB models. In particular astrophysical
results referring to SMBHBs could be significantly biased depending on the chosen GWB
models. Essentially these conclusions are consistent with the model-dependent results presented by
\citet{aaa+23d} and \citet{aaa+23i}. 

%

It is worth noting that the GW analyses carried out by the PTAs makes several simplifying
assumptions about the GWB (Gaussianity, isotropy, and often a power-law spectrum), none of which are
exactly satisfied by a GWB produced by a finite number of supermassive binary black
holes. \citet{bcm+23} studied the impact of these assumptions by generating simulated data sets with
astrophysically realistic GWB and using standard methods to recover their properties.  They found
that the analyses were robust and able to detect the GWB with these simplying assumptions, although
there was a large variance in the significance and recovered parameters.

%
%
%

\subsection{Early Universe: Inflation and Strings}
A number of different mechanisms (many of which relate to cosmic strings) can create a GWB at the
earliest times of the Universe's evolution, as reviewed recently by \citet{lv22}. Under the
assumption that the signals described in Section~\ref{ssec:HD} arise from that era, these
theoretical predictions can be tested. Based on the most recent NANOGrav results, \citet{aaa+23g}
investigated the consistency of the observations with various cosmological interpretations, namely
inflation, scalar-induced GWs, first-order phase transitions, cosmic strings and domain walls. They find
that when performing a simple phenomenological fit to the observed power spectrum of the common signal most cosmological models provide a better match to the data than the GWB from SMBHBs does, with the exception of one model which they rule out, the stable cosmic strings that have a field-theory
origin. However, the reader should be cautioned that the attained evidence for various
  models explaining the observed signal is very weak and is expected to change as more realistic
  simulations of the SMBHB background and improved noise models are incorporated in the analysis. \citet{aaa+23i} analysed the EPTA/InPTA results and come to an overall similar conclusion: both astrophysical and cosmological models can explain the data equally well. Though, one should point out that for most of the signals of cosmological origin the obtained best-fit parameters would be challenging to interpret in the frame of standard inflation.
A similar conclusion based on the NANOGrav results
is reached by \citet{vag23} who points out that an ``extremely blue tensor spectrum'' would be
required to make the observations consistent with an inflationary origin. After analyzing a wide
variety of early-universe models, \citet{vag23} concludes that the ekpyrotic model of the early
universe \citep{kost21} appears the most natural fit to the data, even though these models are
typically expected to generate a GWB with a significantly lower amplitude than the one observed in PTA data. 

%
%

\subsection{Ultra-light Dark Matter}

A number of different models have been proposed that predict an impact of ULDM on PTA data. Specifically, \citet{kr14} developed a description in which ULDM interacts with
normal matter purely gravitationally. This would introduce oscillating gravitational potentials
which would affect the propagation of photons through space and hence affect pulsar timing
results. This model was the basis for the limits placed by \citet{pzl+18}, based on PPTA data. They
improved previously derived limits by a factor of two to five, resulting in a limit on the
dark-matter density of ultralight bosons of: $\rho < 6\,$GeV/cm$^3$ at 95\% confidence in the mass
range $m\leq 10^{-23}\,$eV. A more constraining result in a slightly more restricted mass range was
recently placed by \citet{sgb+23} based on the EPTA data: $\rho \leq 0.3\,$GeV/cm$^3$ in the mass
range $10^{-24}{\,\rm eV}\leq m\leq10^{-23.4}\,{\rm eV}$, implying that fuzzy dark matter in this mass range is disfavoured by pulsar timing data. The constraints found by \citet{aaa+23g} are very
comparable: $\rho \leq 0.4\,$GeV/cm$^3$ for $m\leq
10^{-23}$\,eV.

Independently, \citet{aaa+23g} used the most recent NANOGrav data to place a limit on ULDM
interacting with the particles of the standard model.
While the physical mechanisms for how the ULDM affects pulsar
timing measurements differ greatly between the two cases, the observational effects are ultimately
highly similar. 
Specifically, \citet{kmt22} do not consider gravitational interaction, but instead consider the case
in which ULDM is directly coupled to 
standard matter through QED or QCD sectors, which causes variations in the moment of inertia of
pulsars or shifts in the terrestrial atomic time standards, which affects the
pulsar timing results.
In particular, the effect of the latter shows a strictly monopolar correlation between the residuals
of different pulsars. Additionally, the background of coupled vector ULDM particles can induce dark
``electric'' fields, which cause periodic displacements between the Earth and any given pulsar \citep{gmr16,xxz+22}. 
Through these mechanisms, \citet{aaa+23g} set stringent upper limits on the coupling constant
between ULDM and standard model particles which are competitive with (and in some cases outperform) laboratory constraints in the mass range $10^{-24}{\,\rm eV}\leq m\leq10^{-20}\,{\rm eV}$ \citep{hga+16, opnet+20, bbm+18}.

%
%

\subsection{Alternative Polarization Modes}
Studying the polarization modes of gravitational waves with LIGO-Virgo-KAGRA (LVK) detectors is
challenging due to the small number of interferometer arms to date, the short duration of most
detected signals and the fact that the design of the LVK detectors is tuned to standard polarization
modes predicted by general relativity \citep{cyc12,plw+20}. It is possible, though, to put stringent
constraints on the presence of alternative polarizations using the current network of detectors
\citep{ligo2021,ligo2021arx}. The limits can be further improved in the presence of an identifiable
electromagnetic counterpart \citep{cyc12} or through observation of strongly lensed GW signals
\citep{m22}. However, in order to fully characterize all six distinct polarization modes that are
predicted by alternative theories of gravity, one needs six linearly independent GW detectors \citep{cih+21}.
For PTAs, the situation is drastically different, primarily due to the large number of Earth-pulsar
detection ``arms''  and there are
realistic prospects to probe the polarization of GWs \citep{ljp08,coty18}. Efforts to place
constraints on GW polarization modes were recently undertaken by \citet{wch22,aaa+23j}, but so far no
distinction can be made between the standard and alternative gravity theories.


\section{Anticipated Improvements}\label{sec:Future}
The scaling laws for PTA sensitivity to a GWB depend on a number of factors related to the PTA's
design and to the detection regime the PTA is in. Since it appears PTAs are approaching a detection,
typically the scaling law is presumed to be the following \citep{sejr13,mtg15}:
\begin{equation}\label{eq:scaling}
  S/N \propto N\sqrt{T}\left(\frac{A\sqrt{C}}{\sigma}\right)^{3/13},
\end{equation}
where $S/N$ is the signal-to-noise ratio of the GWB in the data, $N$ is the number of pulsars in the
array, $T$ is the timing baseline of the PTA data set, $A$ is the dimensionless amplitude of the GWB
at the reference frequency of 1\,yr$^{-1}$, $C$ is the cadence of the observations and $\sigma$ is
the level of the timing precision, typically taken to be the weighted RMS of the timing
residuals. This equation provides a number of clear paths forward to effectively boost the
sensitivity of PTAs to a GWB.

\paragraph{International Collaboration} The foundation of the IPTA in 2008 \citep{man13} was a
logical step towards developing PTA sensitivity, since global combination would immediately increase
$N$ (as all-sky access becomes possible with telescopes in both hemispheres) and $C$. While this
appears straightforward on paper, the IPTA data combinations turned out to be more complicated than
anticipated, although an improvement in GWB sensitivity by a factor of at least two was demonstrated
\citep{vlh+16}. Given the recent results from constituent PTAs, it is anticipated that the next IPTA
data combination \citep{gi23} may provide the first GWB detection with a significance beyond 5\,$\sigma$. 

\paragraph{Continued Observations} In the low-S/N regime, PTA sensitivity scales with $T^{13/3}$
and hence GWB limits increase rapidly \citep{sejr13}. However, as we approach a detection, a
transition occurs and detection significance only increases slowly with observing timespan,
i.e.\ with $T^{1/2}$. Due to the complexities inherent to long data sets \citep[see,
  e.g.,][]{eab+23}, which have additional impacts from timing noise and interstellar dispersion
variations, and which often suffer from inhomogeneities due to changes in instrumentation, longer
timespans in the detection era may provide more challenges than benefits. This is underscored by the
power of highly sensitive telescopes like MeerKAT and FAST, which have PTAs with short timing
baselines which are still competitive with the older PTAs that have far longer timespans
\citep{xcg+23,msb+23}.

\paragraph{New Telescopes} As hinted in the previous paragraph, new, more powerful telescopes like
the MeerKAT telescope array or the Chinese FAST telescope, can significantly improve the timing
precision of MSPs by beating down the white noise. A case in point provides the timing of
PSR~J1933$-$6211, which was timed by both the Parkes telescope \citep{gvo+17} and MeerKAT
\citep{gvf+23} in recent years. The enhanced timing precision provided by MeerKAT enables a lot more
science with a lot less complications \citep[see, e.g.,][]{hkc+22}. The same conclusion is likely to hold true for PTA
research. In addition to MeerKAT \citep{bja+20} and FAST \citep{nlj+11}, in North America the
Deep Synoptic Array \citep[DSA-2000,][]{hrw+19} has PTA science as its third key science objective
and is planning to allocate 25\% of its observing time to it. Finally, construction has commenced on
the Square Kilometre Array (SKA) telescope, which is also expected to host a high-impact PTA program
\citep{jhm+15}. 

\paragraph{Wide-band receivers} A major change in the sensitivity of pulsar timing experiments over the
past several decades, has been the development of wide-band receiver systems. The PPTA has commenced
PTA observations with such a system \citep{cpb+23}, having a bandwidth from 704\,MHz up to
4.032\,GHz \citep{hmd+20}. The key advantages of such systems for PTA work are an increase in
sensitivity (lowering $\sigma$) but also additional sensitivity to frequency-dependent effects,
which may aid noise-modeling and thereby increase sensitivity. Similar plans for wide-band
receivers have been developed for the GBT and the Effelsberg radio telescopes, although PTA
observations with those systems have so far not been published.

\paragraph{Diverse PTAs} So far, most PTAs have concentrated their observing campaigns on the
GHz frequency range. With the upcoming addition of the low-frequency LOFAR \citep{dvt+20}, NenuFAR \citep{bgt+21} and CHIME
\citep{cab+21} radio telescopes, data with additional sensitivity to propagation effects and with
higher cadence will become possible. Furthermore, the recently proposed Gamma-ray PTA \citep{faa+22}
will complement these efforts by providing a timing baseline that is fully unaffected by
interstellar propagation effects. 

\begin{figure}
	\centering 
	\includegraphics[width=0.5\textwidth, angle=0]{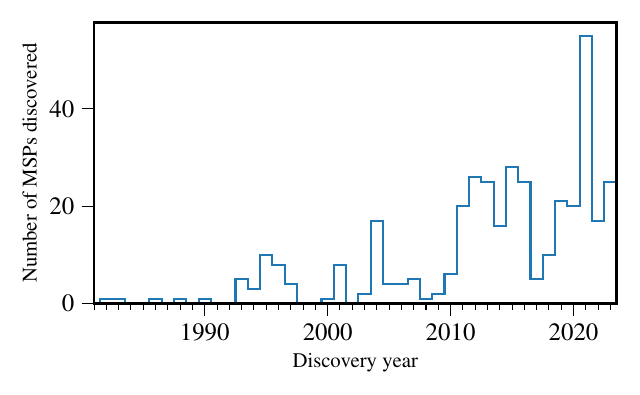}	
	\caption{Number of Galactic disk MSPs discovered as a function of time. Primarily due to the
      surveys with a new generation of highly sensitive telescopes (FAST, MeerKAT), the discovery
      rate in recent years has remained as high as it has ever been. Data from the ATNF pulsar
      catalog, version 2.1.0 \citep{mhth05}. Note that a more up-to-date list of MSP discoveries is
      maintained at \protect\url{https://www.astro.umd.edu/~eferrara/pulsars/GalacticMSPs.txt},
      which presently contains 49 more unpublished MSPs, primarily discovered in the last $\sim5$
      years. For the purpose of this plot, however, we chose to only display published pulsar
      discoveries.}
    %
	\label{fig:MSPHist}%
\end{figure}

\begin{figure}
  \centering
  \includegraphics[width=0.5\textwidth,angle=0]{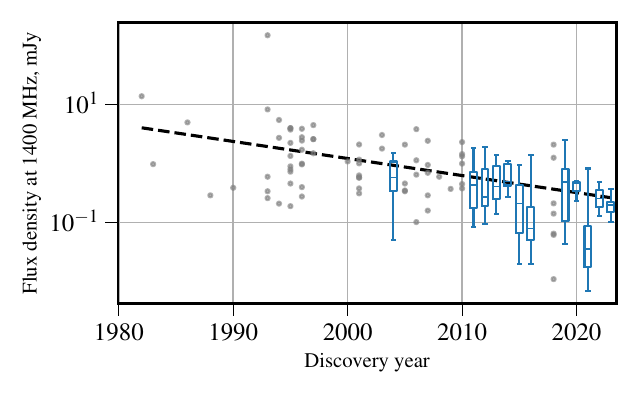}
  \caption{Flux density at 1.4\,GHz observing frequency of disk MSPs as a function of discovery
    year. In years where more than 15 data points are available, box plots are shown instead of the
    individual measurements. The box plots show the median as a horizontal line, the 25-75$^{\rm
      th}$ percentiles as a box and the extreme values as whiskers. As expected for ever more
    complex and ever deeper surveys, the brightness of newly discovered MSPs has consistently
    trended downwards. \citep[Data from ATNF catalogue version 2.1.0,][]{mhth05}}
  \label{fig:MSPFlux}
\end{figure}

\paragraph{More Pulsars} As can be seen in Figure~\ref{fig:MSPHist}, even though the pulsar surveys
from the mid-1990s and early 2000s were considered sufficiently successful to warrant the
commencement of PTA experiments, discovery rates in recent times have consistently been higher than
back then. This bodes well for PTA experiments, not only because the number of MSPs in PTAs can be
continuously increased, thereby enhancing sensitivity to GWs, but also because the best MSPs can be
selected from an ever-growing population. However, part of the reason that so many new MSPs are
being discovered in recent times, is that observing hardware has been upgraded, thereby allowing
fainter sources to be discovered. As shown in Figure~\ref{fig:MSPFlux}, the average flux density of
discovered disk MSPs has persistently decreased, implying that any newly discovered MSP is fainter
and hence on average less useful for high-precision timing efforts. This trend is clearly partly
counter-acted by the increased sensitivity of observing systems and new telescopes, but does clarify
an increased reliance on that additional sensitivity. (We restricted ourselves to disk MSPs, which
were defined as pulsars with spin periods shorter than 30\,ms and spin-down rate below
$10^{-15}$\,s/s and which are not associated with globular clusters. While many MSPs inhabit
globular clusters, due to the high density in the cluster environment, these MSPs undergo
unpredictable accelerations that add strong red noise in their timing, thereby making their
inclusion in PTAs particularly problematic.)


\paragraph{Novel Analysis Methods}
The primary challenge in current PTA analyses is the computational load and processing time,
particularly for the Bayesian GW analysis and noise modeling. Consequently, research is being
undertaken along multiple lines to attempt and improve the processing speed, for example by
optimizing Bayesian samplers \citep{llr+23}, or by applying artificial intelligence methods
\citep{sft+23}. Since both noise modeling and GW analysis typically rely on spectral analysis,
optimized routines for spectral characterization have also recently been proposed by
\citet{ltv23}. (Note that the uneven sampling of PTA data sets is usually not a problem since the
analysis routines Fourier-transform the models rather than the data.)





\section{Discussion and Conclusions}\label{sec:Conc}
PTAs hold the prospect of detecting GWs from a diversity of sources, although most commonly they are
expected to probe a GWB created by multitudes of SMBHBs distributed throughout the Universe. In
recent years, multiple indications have been presented that make a convincing case supporting the
expectation that a full detection of GWs by PTAs is imminent. This would open the nHz-frequency part
of the GW spectrum and provide GW observations that are highly complementary to those provided by
the LIGO/Virgo/KAGRA collaboration. PTAs are accelerating their progress towards detection through a
diverse combination of efforts, not in the least through international collaboration, but also
through continued hardware improvements and improvements in data-analysis techniques and by adding
complementary observations to their analyses. Limits and preliminary results that hint at a future
detection are already being used to constrain a diverse set of cosmological, gravitational and
astrophysical models for GWBs and while the results are generally not fully consistent with standard
model predictions, preferred models can still be tuned to match observations. This suggests that
even the first 5-$\sigma$ detection may already provide significant astrophysical insights in
processes that have so far been the monopoly of N-body simulations. 

\section*{Acknowledgements}
The authors wish to express their gratitude to the lively, passionate and highly diverse community
that drives the exciting research summarized in this paper. We are particularly grateful to the
technical staff and operators at the many radio telescopes across the world, who keep their
instruments in top shape and continue to provide the highest quality data that are required for the
GW detection we are all jointly moving towards. The authors particularly wish to thank the two
referees and the following colleagues:
Maura McLaughlin,
Golam Shaifullah,
Deborah Good,
Boris Goncharov,
Michael Lam,
Krishnakumar M.\ A.,
William Lamb,
Jackson Taylor and
Jeremy Baier for useful discussions, comments and suggestions, which have significantly improved
this paper. 
JPWV acknowledges support through the Space Research Initiative (SRI, award number AWD00005725), managed by the Florida Space Institute (FSI) and from NSF AccelNet award No.\ 2114721. 
SJV acknowledges support from NSF award No.\ 2309246 and NSF Physics Frontiers Center award No.\ 2020265.
NKP funded by the Deutsche
Forschungsgemeinschaft (DFG, German Research Foundation) --
Projektnummer PO 2758/1--1, through the Walter--Benjamin
programme.
Parts of this research were conducted by the Australian Research Council Centre of Excellence for Gravitational Wave Discovery (OzGrav), through project number CE170100004.

\appendix

\section{List of PTA Pulsars}
\label{app:PSRs}
\onecolumn
\begin{table}
  \begin{tabular}{lrrrrccccccc}
    \hline
    Name & Period & DM & $P_{\rm b}$ & $S_{1400}$  & \multicolumn{7}{c}{Observed by}\\
    (J2000) & (ms) & (pc/cm$^3$) & (days) & (mJy) & EPTA & InPTA & NANOGrav & PPTA & CPTA & MPTA
    & LOFAR \\
    \hline
J0023+0923   &  3.05 & 14.3 & 0.1 & 0.7 & - & - & V & - & V & - & -  \\
J0030+0451   &  4.87 &  4.3 & --  & 1.1 & V & - & V & V & V & V & V  \\
J0034$-$0534 &  1.88 & 13.8 & 1.6 & 0.2 & - & - & - & - & V & - & V  \\
J0125$-$2327 &  3.68 &  9.6 & --  & 2.5 & - & - & - & V & - & V & -  \\
J0154+1833   &  2.37 & 19.8 & --  & 0.1 & - & - & - & - & V & - & -  \\
J0218+4232   &  2.32 & 61.3 & 2.0 & 0.9 & - & - & - & - & V & - & V  \\
J0340+4130   &  3.30 & 49.6 & --  & 0.3 & - & - & V & - & V & - & -  \\
J0406+3039   &  2.61 & 49.4 & 7.0 & --  & - & - & V & - & V & - & -  \\
J0407+1607   & 25.70 & 35.6 &669.1& 0.4 & - & - & - & - & - & - & V  \\
J0437$-$4715 &  5.76 &  2.6 & 5.7 &150.2& - & V & V & V & - & V & -  \\
J0509+0856   &  4.06 & 38.3 & 4.9 & 1.5 & - & - & V & - & V & - & -  \\
J0557+1551   &  2.56 &102.6 & 4.8 & --  & - & - & V & - & - & - & -  \\
J0605+3757   &  2.73 & 20.9 &55.7 & --  & - & - & V & - & V & - & -  \\
J0610$-$2100 &  3.86 & 60.7 & 0.3 & 0.7 & - & - & V & - & - & V & -  \\
J0613$-$0200 &  3.06 & 38.8 & 1.2 & 2.3 & V & V & V & V & V & V & -  \\
J0614$-$3329 &  3.15 & 37.1 &53.6 & 0.7 & - & - & V & V & - & V & -  \\
J0621+1002   & 28.85 & 36.6 & 8.3 & 1.7 & - & - & - & - & V & - & V  \\
J0636$-$3044 &  3.95 & 15.5 & --  & 1.5 & - & - & - & - & - & V & -  \\
J0636+5128   &  2.87 & 11.1 & 0.1 & 1.0 & - & - & V & - & V & - & -  \\
J0645+5158   &  8.85 & 18.2 & --  & 0.3 & - & - & V & - & V & - & V  \\
J0709+0458   & 34.43 & 44.3 & 4.4 & 0.3 & - & - & V & - & - & - & -  \\
J0711$-$6830 &  5.49 & 18.4 & --  & 2.6 & - & - & - & V & - & V & -  \\
J0732+2314   &  4.09 & 44.7 &30.2 & 0.7 & - & - & - & - & V & - & -  \\
J0740+6620   &  2.89 & 15.0 & 4.8 & 1.1 & - & - & V & - & - & - & V  \\
J0751+1807   &  3.48 & 30.2 & 0.3 & 1.4 & V & V & - & - & V & - & V  \\
J0824+0028   &  9.86 & 34.5 &23.2 & 0.5 & - & - & - & - & V & - & -  \\
J0900$-$3144 & 11.11 & 75.7 &18.7 & 3.8 & V & - & - & V & - & V & -  \\
J0931$-$1902 &  4.64 & 41.5 & --  & 0.5 & - & - & V & - & - & V & -  \\
J0952$-$0607 &  1.41 & 22.4 & 0.3 & --  & - & - & - & - & - & - & V  \\
J0955$-$6150 &  2.00 &160.9 &24.6 & 0.6 & - & - & - & - & - & V & -  \\
J1012$-$4235 &  3.10 & 71.7 &38.0 & 0.3 & - & - & V & - & - & V & -  \\
J1012+5307   &  5.26 &  9.0 & 0.6 & 3.8 & V & V & V & - & V & - & V  \\
J1017$-$7156 &  2.34 & 94.2 & 6.5 & 1.1 & - & - & - & V & - & V & -  \\
J1022+1001   & 16.45 & 10.3 & 7.8 & 3.9 & V & V & V & V & - & V & V  \\
J1024$-$0719 &  5.16 &  6.5 & --  & 1.5 & V & - & V & V & V & V & V  \\
J1036$-$8317 &  3.41 & 27.1 & 0.3 & 0.5 & - & - & - & - & - & V & -  \\
J1045$-$4509 &  7.47 & 58.1 & 4.1 & 2.7 & - & - & - & V & - & V & -  \\
J1101$-$6424 &  5.11 &207.4 & 9.6 & 0.3 & - & - & - & - & - & V & -  \\
J1103$-$5403 &  3.39 &103.9 & --  & 0.4 & - & - & - & - & - & V & -  \\
J1125$-$5825 &  3.10 &124.8 &76.4 & 1.0 & - & - & - & - & - & V & -  \\
J1125$-$6014 &  2.63 & 52.9 & 8.8 & 1.3 & - & - & - & V & - & V & -  \\
J1125+7819   &  4.20 & 12.0 &15.4 & 1.1 & - & - & V & - & - & - & V  \\
J1216$-$6410 &  3.54 & 47.4 & 4.0 & 1.2 & - & - & - & - & - & V & -  \\
J1300+1240$^a$& 6.22 & 10.2 &25.3 & 0.4 & - & - & - & - & - & - & V  \\
J1312+0051   &  4.23 & 15.3 &38.5 & 0.2 & - & - & V & - & - & - & -  \\
J1327$-$0755 &  2.68 & 27.9 & 8.4 & 0.2 & - & - & - & - & - & V & -  \\
J1327+3423   & 41.51 &  4.2 & --  & --  & - & - & - & - & V & - & -  \\
J1400$-$1431 &  3.08 &  4.9 & 9.5 & 0.2 & - & - & - & - & - & - & V  \\
\hline
\end{tabular}
  \caption{MSPs that are being timed as part of a PTA experiment. Listed are the J2000 name of the
    pulsar, its pulse period, interstellar dispersion measure, orbital period (if any), measured
    flux density at 1.4\,GHz and which PTAs it is observed by ('V' indicates the pulsar is part of
    the array, '-' indicates it is not). Values were taken from the ATNF Pulsar Catalogue \citep{mhth05}, \citet{aaa+23a} or from \citet{flm+23} in the case of PSR~J1327+3423. The source lists from the various observing campaigns were taken from the references cited in Section~\ref{ssec:Data} $^a$: PSR~J1300+1240 is also referred to as PSR~B1257+12.}
 \label{tab:PSRs}
 \end{table}

\begin{table}
  \begin{tabular}{lrrrrccccccc}
    \hline
    Name & Period & DM & $P_{\rm b}$ & $S_{1400}$  & \multicolumn{7}{c}{Observed by}\\
    (J2000) & (ms) & (pc/cm$^3$) & (days) & (mJy) & EPTA & InPTA & NANOGrav & PPTA & CPTA & MPTA
    & LOFAR \\
    \hline
J1421$-$4409 &  6.39 &  54.6 & 30.7 & 1.3 & - & - & - & - & - & V & -  \\
J1431$-$5740 &  4.11 & 131.4 &  2.7 & 0.4 & - & - & - & - & - & V & -  \\
J1435$-$6100 &  9.35 & 113.8 &  1.4 & 0.3 & - & - & - & - & - & V & -  \\
J1446$-$4701 &  2.20 &  55.8 &  0.3 & 0.5 & - & - & - & V & - & V & -  \\
J1453+1902   &  5.79 &  14.1 &  --  & 0.2 & - & - & V & - & V & - & -  \\
J1455$-$3330 &  7.99 &  13.6 & 76.2 & 0.7 & V & - & V & - & - & V & -  \\
J1525$-$5545 & 11.36 & 127.0 &  1.0 & 0.4 & - & - & - & - & - & V & -  \\
J1543$-$5149 &  2.06 &  51.0 &  8.1 & 0.8 & - & - & - & - & - & V & -  \\
J1544+4937   &  2.16 &  23.2 &  0.1 & --  & - & - & - & - & - & - & V  \\
J1545$-$4550 &  3.58 &  68.4 &  6.2 & 1.1 & - & - & - & V & - & V & -  \\
J1547$-$5709 &  4.29 &  95.7 &  3.1 & 0.3 & - & - & - & - & - & V & -  \\
J1552+5437   &  2.43 &  22.9 &  --  & --  & - & - & - & - & - & - & V  \\
J1600$-$3053 &  3.60 &  52.3 & 14.3 & 2.4 & V & V & V & V & - & V & -  \\
J1603$-$7202 & 14.84 &  38.0 &  6.3 & 2.5 & - & - & - & V & - & V & -  \\
J1614$-$2230 &  3.15 &  34.5 &  8.7 & 1.1 & - & - & V & - & - & V & -  \\
J1629$-$6902 &  6.00 &  29.5 &  --  & 1.0 & - & - & - & - & - & V & -  \\
J1630+3734   &  3.32 &  14.2 & 12.5 & --  & - & - & V & - & V & - & -  \\
J1640+2224   &  3.16 &  18.4 &175.5 & 0.5 & V & - & V & - & V & - & V  \\
J1643$-$1224 &  4.62 &  62.4 &147.0 & 3.8 & - & V & V & V & V & V & -  \\
J1652$-$4838 &  3.79 & 188.2 &  --  & 0.9 & - & - & - & - & - & V & -  \\
J1653$-$2054 &  4.13 &  56.5 &  1.2 & 0.6 & - & - & - & - & - & V & -  \\
J1658+3630   & 33.03 &   3.0 &  3.0 & --  & - & - & - & - & - & - & V  \\
J1658$-$5324 &  2.44 &  30.8 &  --  & 0.4 & - & - & - & - & - & V & -  \\
J1705$-$1903 &  2.48 &  57.5 &  --  & 0.6 & - & - & V & - & - & V & -  \\
J1708$-$3506 &  4.51 & 146.8 &149.1 & 1.5 & - & - & - & - & - & V & -  \\
J1710+4923   &  3.22 &   7.1 &  --  & --  & - & - & - & - & V & - & -  \\
J1713+0747   &  4.57 &  16.0 & 67.8 & 8.3 & V & V & V & V & V & V & V  \\
J1719$-$1438 &  5.79 &  36.8 &  0.1 & 0.4 & - & - & V & - & - & V & -  \\
J1721$-$2457 &  3.50 &  48.2 &  --  & 0.6 & - & - & - & - & - & V & -  \\
J1730$-$2304 &  8.12 &   9.6 &  --  & 4.0 & V & - & V & V & - & V & V  \\
J1732$-$5049 &  5.31 &  56.8 &  5.3 & 2.1 & - & - & - & - & - & V & -  \\
J1737$-$0811 &  4.18 &  55.3 & 79.5 & 1.1 & - & - & - & - & - & V & -  \\
J1738+0333   &  5.85 &  33.8 &  0.4 & 0.3 & V & - & V & - & V & - & V  \\
J1741+1351   &  3.75 &  24.2 & 16.3 & 0.3 & - & - & V & V & V & - & -  \\
J1744$-$1134 &  4.08 &   3.1 &  --  & 2.6 & V & V & V & V & V & V & V  \\
J1745+1017   &  2.65 &  24.0 &  0.7 & 0.5 & - & - & V & - & V & - & -  \\
J1747$-$4036 &  1.65 & 152.9 &  --  & 1.5 & - & - & V & - & - & V & -  \\
J1751$-$2857 &  3.92 &  42.8 &110.7 & 0.5 & V & - & V & - & - & V & -  \\
J1756$-$2251 & 28.46 & 121.2 &  0.3 & 1.1 & - & - & - & - & - & V & -  \\
J1757$-$5322 &  8.87 &  30.8 &  0.5 & 1.2 & - & - & - & - & - & V & -  \\
J1801$-$1417 &  3.63 &  57.3 &  --  & 1.5 & V & - & - & - & - & V & -  \\
J1802$-$2124 & 12.65 & 149.6 &  0.7 & 0.7 & - & - & V & - & - & V & -  \\
J1804$-$2717 &  9.34 &  24.7 & 11.1 & 0.4 & V & - & - & - & - & - & -  \\
J1811$-$2405 &  2.66 &  60.6 &  6.3 & 1.3 & - & - & V & - & - & V & -  \\
J1824$-$2452 &  3.05 & 119.9 &  --  & 2.3 & - & - & - & V & - & - & -  \\
J1825$-$0319 &  4.55 & 119.6 & 52.6 & 0.2 & - & - & - & - & - & V & -  \\
J1832$-$0836 &  2.72 &  28.2 &  --  & 0.9 & - & - & V & V & V & V & -  \\
J1843$-$1113 &  1.85 &  60.0 &  --  & 0.1 & V & - & V & - & V & V & -  \\
J1853+1303   &  4.09 &  30.6 &115.7 & 0.5 & - & - & V & - & V & - & V  \\
J1857+0943$^b$& 5.36 &  13.3 & 12.3 & 5.0 & V & V & V & V & V & - & V  \\
\hline
\end{tabular}
  \caption{MSPs that are being timed as part of a PTA experiment. Continued from
    Table~\ref{tab:PSRs}. $^b$: PSR~J1857+0943 is also referred to as PSR~B1855+09.}
 \label{tab:PSRs2}
 \end{table}

\begin{table}
  \begin{tabular}{lrrrrccccccc}
    \hline
    Name & Period & DM & $P_{\rm b}$ & $S_{1400}$ & \multicolumn{7}{c}{Observed by}\\
    (J2000) & (ms) & (pc/cm$^3$) & (days) & (mJy) & EPTA & InPTA & NANOGrav & PPTA & CPTA & MPTA
    & LOFAR \\
    \hline
J1902$-$5105 &  1.74 &  36.3 &  2.0 & 1.0 & - & - & - & V & - & V & -  \\
J1903+0327   &  2.15 & 297.5 & 95.2 & 0.6 & - & - & V & - & V & - & -  \\
J1903$-$7051 &  3.60 &  19.7 & 11.1 & 1.0 & - & - & - & - & - & V & -  \\
J1909$-$3744 &  2.95 &  10.4 &  1.5 & 1.8 & V & V & V & V & - & V & -  \\
J1910+1256   &  4.98 &  38.1 & 58.5 & 0.7 & V & - & V & - & V & - & -  \\
J1911$-$1114 &  3.63 &  31.0 &  2.7 & 1.0 & - & - & - & - & V & - & V  \\
J1911+1347   &  4.63 &  31.0 &  --  & 0.9 & V & - & V & - & V & - & -  \\
J1918$-$0642 &  7.65 &  26.6 & 10.9 & 0.6 & V & - & V & - & V & V & V  \\
J1923+2515   &  3.79 &  18.9 &  --  & 0.2 & - & - & V & - & V & - & V  \\
J1933$-$6211 &  3.54 &  11.5 & 12.8 & 1.0 & - & - & - & V & - & V & -  \\
J1939+2134$^c$& 1.56 &  71.0 &  --  &13.9 & - & V & V & V & - & - & -  \\
J1944+0907   &  5.19 &  24.4 &  --  & 2.1 & - & - & V & - & V & - & V  \\
J1946+3417   &  3.17 & 110.2 & 27.0 & 0.9 & - & - & V & - & V & - & -  \\
J1946$-$5403 &  2.71 &  23.7 &  0.1 & 0.4 & - & - & - & - & - & V & -  \\
J1955+2908$^d$& 6.13 & 104.5 &117.3 & 1.0 & - & - & V & - & V & - & V  \\
J2010$-$1323 &  5.22 &  22.2 &  --  & 0.7 & - & - & V & - & V & V & -  \\
J2017+0603   &  2.90 &  23.9 &  2.2 & 0.2 & - & - & V & - & V & - & -  \\
J2019+2425   &  3.94 &  17.2 & 76.5 & 0.3 & - & - & - & - & V & - & -  \\
J2022+2534   &  2.65 &  53.7 &  --  &  -- & - & - & - & - & V & - & -  \\
J2033+1734   &  5.95 &  25.1 & 56.3 & 0.3 & - & - & V & - & V & - & -  \\
J2039$-$3616 &  3.28 &  24.0 &  5.8 & 0.5 & - & - & - & - & - & V & -  \\
J2043+1711   &  2.38 &  20.7 &  1.5 & 0.1 & - & - & V & - & V & - & V  \\
J2051$-$0827 &  4.51 &  20.7 &  0.1 & 2.8 & - & - & - & - & - & - & V  \\
J2124$-$3358 &  4.93 &   4.6 &  --  & 4.5 & V & V & V & V & - & V & -  \\
J2129$-$5721 &  3.73 &  31.8 &  6.6 & 1.0 & - & - & - & V & - & V & -  \\
J2145$-$0750 & 16.05 &   9.0 &  6.8 & 5.5 & - & V & V & V & V & V & V  \\
J2150$-$0326 &  3.50 &  20.7 &  --  & 0.5 & - & - & - & - & V & V & -  \\
J2214+3000   &  3.12 &  22.5 &  0.4 & 0.5 & - & - & V & - & V & - & -  \\
J2222$-$0137 & 32.82 &   3.3 &  2.4 & 0.9 & - & - & - & - & - & V & V  \\
J2229+2643   &  2.98 &  22.7 & 93.0 & 0.8 & - & - & V & - & V & V & -  \\
J2234+0611   &  3.58 &  10.8 & 32.0 & 0.5 & - & - & V & - & V & - & -  \\
J2234+0944   &  3.63 &  17.8 &  0.4 & 1.9 & - & - & V & - & V & V & -  \\
J2241$-$5236 &  2.19 &  11.4 &  0.1 & 1.8 & - & - & - & V & - & V & -  \\
J2302+4442   &  5.19 &  13.8 &125.9 & 1.4 & - & - & V & - & V & - & V  \\
J2317+1439   &  3.45 &  21.9 &  2.5 & 0.6 & - & - & V & - & V & V & V  \\
J2322+2057   &  4.81 &  13.4 &  --  & 0.3 & V & - & V & - & V & V & -  \\
J2322$-$2650 &  3.46 &   6.1 &  0.3 & 0.2 & - & - & - & - & - & V & -  \\
\hline
\end{tabular}
  \caption{MSPs that are being timed as part of a PTA experiment. Continued from
    Table~\ref{tab:PSRs}. $^c$: PSR~J1939+2134 is also referred to as PSR~B1937+21. $^d$: PSR~J1955+2908
    is also referred to as PSR~B1953+29.}
 \label{tab:PSRs3}
 \end{table}

\newpage
\section{List of PTA Telescopes and Observing Properties}
\label{app:obs}
\begin{table}
\begin{tabular}{llcclrl}
  \hline
  PTA      & Telescope        & $f_{\rm c}$      & BW            & Timespan  & Ref.\\
           &                  & (MHz)            & (MHz)         &           &     \\
  \hline
  EPTA     & Effelsberg (EFF) & 1380, 2487, 4857 & 240, 300, 500 & 1996--    & \citet{eab+23}\\
           & Lovell$^1$ (JBO) & 1532             & 512           & 2009--    &     \\
           & Nan{\c{c}}ay (NRT)& 1484, 1854, 2154, 2539& 512     & 2004--    &     \\
           & Sardinia (SRT)   & 357, 1396        & 105, 500      & 2014--    &     \\
           & Westerbork (WSRT)& 350, 1380, 2273  & 80, 160, 160  & 1999--2015&     \\
           & LEAP             & 1396             & 128           & 2012--    & \citet{bjk+16}\\
           & LOFAR            & 148.9            & 78.1          & 2012--    & \citet{dvt+20}\\
  \\
  InPTA    & GMRT             & 400, 1360        & 200           & 2018--    & \citet{tnr+22}\\

  \\
  NANOGrav$^2$& Arecibo (AO)  & 327, 433, 1456, 2052&50, 24, 600, 460& 2004--2020& \citet{aaa+23a}\\
           & Green Bank (GBT) & 820, 1518        & 180, 640      & 2004--    &     \\
           & VLA              & 1520, 3000       & 800, 1700     & 2015--    &     \\
           & CHIME            & 600              & 400           & 2019--    & \citet{cab+21}\\
  \\
  PPTA     & Parkes (PKS)     & 2368$^3$         & 3328          & 2004--    & \citet{zrk+23}\\
  \\
  MPTA     & MeerKAT          & 1284             & 856           & 2019--    & \citet{msb+23}\\
  \\
  CPTA     & FAST             & 1250             & 500           & 2019--    & \citet{xcg+23}\\
  \hline
 \end{tabular}
\caption{Characteristics of PTA observations. For each PTA the different participating observatories
  are given, along with the current observing frequencies and maximal bandwidths in the different
  observing bands. The time span corresponds to the range of dates over which PTA observations were
  carried out at the observatories, including with older systems that are not detailed in this
  table. Unless otherwise specified, the first reference given for a PTA refers to  all telescopes of that PTA. The first four PTAs in this list (EPTA, InPTA, NANOGrav and PPTA) are all members of the IPTA, which also has a data-sharing agreement with the MPTA. \\ $^1$:
  For PSR~J1713+0747, the Jodrell Bank Observatory also provided data from the Mark II
  telescope. The observation's characteristics are identical to those listed above.\\ $^2$: For NANOGrav, the quoted bandwidths are the \emph{usable} bandwidth, accounting for RFI excision.\\ $^3$: The PPTA presently uses an ultra-wide-bandwidth receiver with continuous frequency coverage from 704 to 4032 MHz. In practice they split up this bandpass in eight sub-bands with center frequencies ranging from 736 to 3312 MHz.}
\label{tab:PTAs}
\end{table}

\newpage
\twocolumn

\bibliographystyle{elsarticle-harv} 






\end{document}